%% file: main.tex
\documentclass[letterpaper]{article} %

\newif\ifExtendedVersion
\ExtendedVersiontrue

\usepackage[]{aaai23}  %
\ifExtendedVersion
\nocopyright
\fi
\usepackage{times}  %
\usepackage{helvet}  %
\usepackage{courier}  %
\usepackage[hyphens]{url}  %
\usepackage{graphicx} %
\usepackage{natbib}  %
\usepackage{caption} %
\usepackage{algorithm}
\usepackage{algorithmic}

\ifExtendedVersion
\usepackage{fancyhdr}
\fi

\usepackage[algo2e]{algorithm2e}

\usepackage{amsmath}
\usepackage{amssymb}
\usepackage{amsthm}
\usepackage{mathtools}

\usepackage{tikz}
\usepackage{pgfplots}
\pgfplotsset{compat=1.17}

\usepackage{verbatim}

\usepackage{colortbl}
\definecolor{LightGray}{rgb}{0.93, 0.93, 0.93}

\usepackage{xspace}
\newcommand{\TTP}[0]{TTP\xspace}
\newcommand{\TTPs}[0]{TTPs\xspace}

\usepackage[textsize=tiny, textwidth=1.5cm]{todonotes}

\newcommand{\Aron}[1]{} %

\usepackage{soul}

\usepackage[capitalize, noabbrev]{cleveref}

\newcommand{\argmax}[0]{\ensuremath{\operatorname{argmax}}}

\newcommand{\setTechniques}[0]{\ensuremath{\mathcal{A}}}
\newcommand{\setIncidents}[0]{\ensuremath{\mathcal{I}}}

\newcommand{\varIncident}[0]{\ensuremath{I}}
\newcommand{\varPriorIncident}[0]{\ensuremath{\hat{I}}}
\newcommand{\varTechnique}[0]{\ensuremath{a}}

\title{Principled Data-Driven Decision Support for Cyber-Forensic Investigations}

\author{Soodeh Atefi \textsuperscript{\rm 1}, Sakshyam Panda \textsuperscript{\rm 2}, Emmanouil Panaousis \textsuperscript{\rm 2}, Aron Laszka \textsuperscript{\rm 3}}
\affiliations{
 \textsuperscript{\rm 1} University of Houston,\\
 \textsuperscript{\rm 2} University of Greenwich,\\
 \textsuperscript{\rm 3} Pennsylvania State University
}

\begin{document}

\ifExtendedVersion
\pagestyle{fancy}
\renewcommand{\headrulewidth}{0pt}
\lhead{Published in the proceedings of the 37th AAAI Conference on Artificial Intelligence (AAAI 2023).}
\rhead{}
\cfoot{\thepage}
\setlength{\headsep}{2.5em}
\setlength{\footskip}{3.5em}
\setlength{\headheight}{0em}
\fi

\maketitle

\begin{abstract}
In the wake of a cybersecurity incident, it is crucial to promptly discover how the threat actors breached security in order to assess the impact of the incident and to develop and deploy countermeasures that can protect against further attacks.
To this end, defenders can launch a cyber-forensic investigation, which discovers the techniques that the threat actors used in the incident.
A fundamental challenge in such an investigation is prioritizing the investigation of particular techniques since the investigation of each technique requires time and effort, but forensic analysts cannot know which ones were actually used before investigating them.
To ensure prompt discovery, it is imperative to provide decision support that can help forensic analysts with this prioritization.
A recent study demonstrated that data-driven decision support, based on a dataset of prior incidents, can provide state-of-the-art prioritization. 
However, this data-driven approach, called DISCLOSE, is based on a heuristic that utilizes only a subset of the available information and does not approximate optimal decisions.
To improve upon this heuristic, we introduce a principled approach for data-driven decision support for cyber-forensic investigations.
We formulate the decision-support problem using a Markov decision process, whose states represent the states of a forensic investigation. 
To solve the decision problem, we propose a Monte Carlo tree search based method, which relies on a $k$-NN regression over prior incidents to estimate state-transition probabilities.
We evaluate our proposed approach on multiple versions of the MITRE ATT\&CK dataset, which
is a knowledge base of adversarial techniques and tactics based on real-world cyber incidents, and demonstrate that our approach outperforms DISCLOSE in terms of techniques discovered per effort spent.
\end{abstract}

\input{intro}

\ifExtendedVersion
The remainder of this paper is organized as follows.
In \cref{sec:model}, we model cyber-forensic investigations as Markov decision processes and formulate the problem of cyber-forensic decision support.
In \cref{sec:approach}, we describe our computational approach, which is based on $k$-NN regression and Monte Carlo tree search. 
In \cref{sec:results}, we evaluate our approach on datasets of real-world incidents, demonstrating that it outperforms the state-of-the-art approach.
In \cref{sec:related}, we give a brief overview of related work.
Finally, in \cref{sec:concl}, we provide concluding remarks.
\fi

\input{model}

\input{approach}

\input{results}

\input{related}
\input{concl}

\section*{Acknowledgements}
This material is based upon work sponsored by the National Science Foundation under Grant No.\ CNS-1850510.
Any opinions, findings, and conclusions or recommendations expressed in this material are those of the authors and do not necessarily reflect the views of the National Science Foundation.
We thank the anonymous reviewers for their valuable feedback and suggestions.

\bibliography{main}

\clearpage
\appendix

\ifExtendedVersion
\input{appendix}

\fi

\end{document}

%% file: intro.tex
\section{Introduction}
\label{sec:intro}

Cybersecurity is an ever growing concern for businesses and individuals alike.
While it is not always possible to prevent cybersecurity incidents, one should always strive to mitigate them promptly and to learn from them as much as possible.
Thus, it is imperative in the aftermath of an incident to promptly discover the techniques that the threat actors used to breach security.
By discovering how threat actors operate, we can %
learn to develop more effective cyber defences and detect---sufficiently early---cyber-attacks that bypass these defences.
To this end, victims of cyber incidents can launch \emph{cyber-forensic investigations}.

Since we do not know in advance which adversarial techniques the threat actors used, we have to prioritize the investigation of these techniques under \emph{uncertainty}, not knowing if the investigation of a technique will waste precious time or reveal crucial information.
In prior work, Nisioti et al.~\shortcite{nisioti2021data} demonstrated that one can effectively prioritize the investigation of techniques based on datasets of prior incidents, exploiting the fact that most threat actors follow common tactics.
However, this existing data-driven approach is based on a heuristic that does not approximate optimal decisions, regardless of the size of the dataset or the available computational power.

To address this limitation, we propose a principled data-driven approach for prioritization.
We introduce a novel model of cyber-forensic investigations based on Markov decision processes, which enables us to formally define the prioritization problem.
Note that this problem is inherently challenging for multiple reasons.
First, datasets of prior incidents are limited in size as many businesses are reluctant to share detailed information about their security.
Second, although threat actors follow common tactics, it is inherently difficult to predict which particular techniques were used in a particular incident.

In light of these challenges, we propose a computational approach for decision support that combines a non-parametric machine-learning model, based on $k$-nearest neighbor, with a Monte Carlo tree search.
By working directly with the data instead of training a parametric model, we get as much ``mileage'' out of the limited data as possible.

%% file: model.tex
\section{Model and Problem Formulation}
\label{sec:model}

\ifExtendedVersion
To formally define the problem of providing optimal decision support for cyber forensics, we first introduce fundamental assumptions and notations for the key elements of cyber-forensic investigations (\cref{sec:model-forensics}). Then, building on these assumptions and notations, we model investigations as Markov decision processes (\cref{sec:model-MDP}), formulating our problem as optimizing decisions within a process (\cref{sec:problem-formulation}).
For a list of symbols used in our model, %
we refer the reader to \cref{table:symbols}.
\fi

\subsection{Adversarial Techniques and Cyber Forensics} %
\label{sec:model-forensics}

\paragraph{Adversarial Techniques}
We consider a forensic investigation whose objective is to discover---as soon as possible---what \emph{techniques} the threat actors used in the incident that is under investigation.
The motivation for discovering how the threat actors breached security is to learn from the incident and to prevent or mitigate future breaches.
We let $\setTechniques$ denote the \emph{set of all techniques} that threat actors may have used in an incident (e.g., DLL search order hijacking, spearphishing attachment, drive-by compromise).
In practice, the set $\setTechniques$ can be taken from a well-known knowledge base, such as MITRE ATT\&CK~\cite{barnum2012standardizing}, which provides a comprehensive list of common adversarial~techniques.

\paragraph{Incident}
Next, we let $\varIncident_Y \subseteq \setTechniques$ denote the \emph{set of techniques that were used} by threat actors in incident $I$; \Aron{we use this only once? omit?}and for ease of presentation, we let $\varIncident_N = \setTechniques \setminus \varIncident_Y$ denote the \emph{set of techniques that were not used} in incident~$I$. 
A key aspect of cyber-forensic investigations is that $\varIncident_Y$ is not known by the forensic experts at the beginning of the investigation.
Since an investigation is typically launched when a breach is detected, forensic experts might know some elements of~$\varIncident_Y$ (e.g., techniques that triggered an intrusion detection system, leading to the detection of the breach); however, discovering set $\varIncident_Y$ is the very objective of the investigation.
To capture this uncertainty that forensic experts face, we model the set of techniques~$\varIncident_Y$ as a \emph{random variable}, whose realization is unknown at the beginning of the investigation.
The distribution of this random variable represents the threat actors' tactics, that is, how likely they are to use certain subsets of techniques in a cyber attack.

\ifExtendedVersion
\input{tables/symbols}

\fi

\paragraph{Prior Incidents}
In practice, we can estimate the distribution of $\varIncident_Y$ from prior incidents, which have already been investigated.
Note that for this estimation, we can use a public repository of prior incidents, which were perpetrated by other threat actors against other targets (e.g., MITRE Cyber Threat Intelligence Repository from~\citeauthor{mitre_cti}).
While there are differences between how different threat actors operate, most threat actors do follow common tactics.
Therefore, subsets of techniques that were frequently used in prior incidents are likely to have been used in the incident that we are currently investigating.
We let $\setIncidents$ denote the set of prior incidents; and we assume that for each prior incident $\varPriorIncident \in \setIncidents$, we know the set $\varPriorIncident_Y \subseteq \setTechniques$ %
and that $\varPriorIncident_Y$ was drawn from the same distribution as $\varIncident_Y$.

\paragraph{Cyber-forensic Investigation}
During the investigation of incident $I$, forensic experts discover the realization of $\varIncident_Y$ step-by-step.
In each step, the experts choose and investigate a technique $\varTechnique \in \setTechniques$ that they have not investigated yet, and they discover whether or not the threat actors used technique~$\varTechnique$.
\Aron{we need to motivate this assumption better (or perhaps not draw attention to it?)}Note that we assume this discovery to be perfect (i.e., if a technique is investigated, experts correctly learn whether it was used or not); our model and computational approach could be generalized in a straightforward manner to account for false negatives and positives in discovery, but this is not the focus of our work.

\paragraph{Costs and Benefits of Forensic Discovery}
To investigate whether a particular technique was used by the threat actors, forensic experts have to spend time, effort, resources, etc. 
We let constant $C_\varTechnique$ denote the \emph{cost of investigating technique}
$\varTechnique \in \setTechniques$, which represents the time, effort, resources, etc. spent.
In practice, we can estimate these costs based on domain experts' knowledge~\cite{nisioti2021data}.
On the other hand, discovering that a technique was used by the threat actors yields benefit since it provides information that we can use to prevent or mitigate future breaches.
We let constant $B_\varTechnique$ denote the \emph{benefit obtained from discovering} that technique $\varTechnique \in \setTechniques$ was used in the incident. 
\Aron{let's be a bit more specific, how did we estimate this for experiments?}In practice, we can estimate these benefit values based on the impacts of using various techniques, which we can take from well-known knowledge bases (e.g., MITRE ATT\&CK~\cite{barnum2012standardizing}).
Some experts may inherently prefer investigating certain actions over others, e.g., because certain actions provide crucial information on how the adversary breached the system. All such preferences can be captured by the benefit values. %

\subsection{Markov Decision Process}
\label{sec:model-MDP}

Based on the assumptions and notations introduced in the preceding subsection, we now model the cyber-forensic investigation of an incident as a \emph{Markov decision process} (MDP)~\cite{puterman2014markov}, whose steps correspond to the step-by-step investigation of the adversarial techniques.
Modeling the cyber-forensic investigation as an MDP provides the foundation for formulating our decision-support problem in the next subsection.
To define an MDP, we have to specify the \emph{state space} of the process, the set of \emph{actions} that can be taken in each state, the \emph{transition probabilities} between subsequent states, and the \emph{immediate reward} for taking a particular action in a particular state.

\paragraph{State Space}
At any step during the investigation of an incident, the forensic experts have already investigated some techniques, while other techniques remain yet to be investigated.
Further, for each technique that the experts have already investigated, they have  discovered whether or not the threat actors used the technique in the incident.
To formalize the state of the investigation process, we let $Y_t \subseteq \setTechniques$ denote the \emph{set of techniques that have been investigated by step $t$ and were used by the threat actors}, and we let $N_t \subseteq \setTechniques$ denote the \emph{set of techniques that have been investigated by step $t$ but were not used by threat actors}.
Note that  by definition, $Y_t \subseteq \varIncident_Y$ and $Y_t \cap N_t = \emptyset$ for every $t$.
At the beginning of the investigation, before the first step, the process is in an \emph{initial state} $\langle Y_0, N_0 \rangle$.
In practice, $Y_0$ can be the set of techniques that triggered an intrusion detection system, leading to the detection of the incident and the launch of the investigation.

\paragraph{Action Space}
At each step $t$, the forensic experts have to choose a technique that they have not investigated yet, and investigate it.
To formalize this decision, we let the set of actions that can be taken in a state correspond to the set of techniques that have not yet been investigated: %
in state $\langle Y_t, N_t \rangle$, the \emph{set of actions} is the set of techniques $\setTechniques \setminus \left( Y_t \cup N_t \right)$.

\paragraph{Transition Probabilities}
Recall that forensic experts do not know in advance (i.e., before investigating) which techniques were used by the threat actors, and we capture this uncertainty by assuming that $\varIncident_Y$ is a random variable whose realization is unknown at the beginning of the investigation.
When experts investigate a technique $\varTechnique$, they discover whether technique $\varTechnique$ was used or not (i.e., if technique $\varTechnique$ is in the realization of $\varIncident_Y$).
Hence, if action $\varTechnique \in \setTechniques \setminus \left( Y_t \cup N_t \right)$ is taken in state $\langle Y_t, N_t\rangle$, then the process transitions either to state $\langle Y_{t+1}, N_{t+1} \rangle = \langle Y_t \cup \{ \varTechnique \}, N_t \rangle$ (if technique $\varTechnique$ was used) or to state $\langle Y_{t+1}, N_{t+1} \rangle = \langle Y_t, N_t \cup \{ \varTechnique \} \rangle$ (if technique $\varTechnique$ was not used). 
The \emph{probabilities of these transitions} are
\begin{align}
& \Pr\left[ \langle Y_{t+1}, N_{t+1} \rangle = \langle Y_t \cup \{ \varTechnique \}, N_t \rangle \right] \nonumber \\
& ~~~~= 
\Pr\left[a \in \varIncident_Y \,|\, Y_t \subseteq \varIncident_Y \wedge N_t \cap \varIncident_Y = \emptyset \right] ~~~~~~~~
\label{eq:trans_prob1}
\end{align}
and 
\begin{align}
& \Pr\left[ \langle Y_{t+1}, N_{t+1} \rangle = \langle Y_t, N_t \cup \{ \varTechnique \} \rangle \right] \nonumber \\
& ~~~~= 
\Pr\left[a \not\in \varIncident_Y \,|\, Y_t \subseteq \varIncident_Y \wedge N_t \cap \varIncident_Y = \emptyset \right] \nonumber \\
& ~~~~= 
1 - \Pr\left[a \in \varIncident_Y \,|\, Y_t \subseteq \varIncident_Y \wedge N_t \cap \varIncident_Y = \emptyset \right] .
\label{eq:trans_prob2}
\end{align}
For ease of notation, we will let $\Pr[a \,|\, Y_t, N_t]$ denote the first probability (i.e., probability that the threat actors used technique $\varTechnique$ given that the state is $\langle Y_t, N_t \rangle$).
In practice, we have to estimate these conditional probabilities based on the set of prior incidents $\setIncidents$.

\paragraph{Rewards}
We formulate rewards to capture the benefits of the cyber-forensic investigation.
The experts obtain a benefit $B_\varTechnique$ from investigating technique $\varTechnique$ only if technique~$\varTechnique$ was used in the incident. 
Hence, if the process transitions from state $\langle Y_t, N_t \rangle$ to state $\langle Y_{t+1}, N_{t+1} \rangle = \langle Y_t \cup \{ \varTechnique \}, N_t \rangle$, then we let the \emph{reward for step $t$ be $B_\varTechnique$};
if the process transitions to state $\langle Y_{t+1}, N_{t+1} \rangle = \langle Y_t, N_t \cup \{ \varTechnique \} \rangle$, then we let the \emph{reward for step $t$ be $0$}.

\subsection{Cyber-forensic Decision-Support Problem}
\label{sec:problem-formulation}

We represent the decision-support system as a policy $\pi$, which maps a state $\langle Y_t, N_t\rangle$ to a recommended action $\varTechnique_t \in \setTechniques \setminus \left( Y_t \cup N_t \right)$ in each time step $t$.
Our goal is to provide a \emph{policy that maximizes the expected rewards obtained during the forensic investigation}.
Following \citeauthor{nisioti2021data}~\shortcite{nisioti2021data}, we formulate this objective with a \emph{budget limit $G$} on the total cost~$\sum C_{a_t}$:
\begin{align}
    \max_{\pi} \mathbb{E}_{I_Y} \Bigg[ \sum_{t=0}^{T_\textit{limit}} 1_{\{\varTechnique_t \in \varIncident_Y\}} \cdot B_{\varTechnique_t}  ~\Bigg|~ \varTechnique_t = \pi\left( Y_t, N_t \right)   \Bigg]
    \label{eq:objective}
\end{align}
where $T_\textit{limit} = \max_T \sum_{t=0}^{T} C_{a_t} \leq G$ (i.e., $T_\textit{limit}$ is the last step before the investigation budget $G$ is exhausted), and $C_{a_t}$ is the cost of investigating the action that is chosen in time step $t$.
Note that we focus on this objective because it is practical and enables a fair comparison with DISCLOSE \cite{nisioti2021data}; in \cref{eq:problem_formulation}, we will provide a more conventional formulation with temporal discounting.

In practice, the budget may be flexible, in which case our goal is to provide a policy that attains a good cost-benefit tradeoff. 
We will quantify this tradeoff in our experiments using the area under the cost-benefit curve (i.e., AUC for the expected benefit as a function of the budget limit).

%% file: tables/symbols.tex
\begin{table}[t]
\small
\centering
\renewcommand{\arraystretch}{1.2}
\begin{tabular}{|l|l|}
\hline 
Symbol   & Description   \\    
\hline
\hline
$\setTechniques$ & set of adversarial techniques (actions in MDP) \\
\rowcolor{LightGray} $\setIncidents$ & set of prior cyber incidents (dataset) \\
$\varIncident_Y \subseteq \setTechniques$ & set of techniques used in incident $I$ \\
\rowcolor{LightGray} $\varIncident_N \subseteq \setTechniques$ & set of techniques not used in incident $I$ \\
$C_\varTechnique$ & effort cost of investigating technique $\varTechnique \in \setTechniques$ \\
\rowcolor{LightGray} $B_\varTechnique$ & benefit from discovering technique $\varTechnique \in \setTechniques$ \\
$Y_t \subseteq \setTechniques$ & set of used techniques discovered  by step $t$ \\
\rowcolor{LightGray} $N_t \subseteq \setTechniques$ & set of not-used techniques discovered  by step $t$ \\
$\Pr[\varTechnique | Y_t, N_t]$ & prob.\ that tech.\ $\varTechnique$ was used given state $\langle Y_t, N_t\rangle$ \\
\rowcolor{LightGray} $\gamma \in (0, 1)$ & temporal discount factor \\
\hline
\end{tabular}
\caption{List of Model Symbols}
\label{table:symbols}
\end{table}

%% file: approach.tex
\section{Computational Approach}
\label{sec:approach}
To solve the decision-support problem based on the objective in \cref{eq:objective}, we propose a computational approach based on \emph{Monte Carlo tree search} (MCTS) and $k$\emph{-nearest neighbour} ($k$-NN) algorithms.
Specifically, we implement the policy~$\pi$ as an MCTS algorithm (\cref{subsection:Tree Search}), %
relying on $k$-NN for estimating transition probabilities (\cref{subsection:Estimating Probabilities}).

Note that in recent years, deep reinforcement learning (DRL) algorithms have garnered significant attention from researchers for their applications in cybersecurity decision support (e.g.,~\cite{ganesan2016dynamic,kurt2018online}).
While we could apply DRL algorithms to our problem in principle, they are ill-suited to our problem for multiple reasons.
First, DRL algorithms tend to be sample inefficient (i.e., require a large number of training experiences to learn a performant policy), which poses a significant challenge since it is difficult to collect large datasets of prior incidents with sufficient details. %
Second, the limited size of the dataset enables us to make decisions based directly on the dataset, instead of having to train a parametric machine-learning model.
By working directly with the dataset, decisions can take advantage of all the information in the dataset.

\subsection{Probability Estimation}
\label{subsection:Estimating Probabilities}

In a nutshell, Monte Carlo tree search finds which action to take in a given state by randomly sampling sequences of actions, simulating how they play out starting from the given state, and choosing the action that leads to the highest rewards on average.
We can simulate the cyber-forensic investigation process from any given state using the state-transition probabilities (\cref{eq:trans_prob1,eq:trans_prob2}). 

\paragraph{Empirical Probabilities}
However, since we do not know the underlying probability distribution in practice, we have to estimate the probabilities based on the set of prior incidents $\setIncidents$.
We can estimate the state-transition probability $\Pr[\varTechnique \,|\, Y_t, N_t]$ as the \emph{conditional empirical probability}:
\begin{align}
    \Pr[\varTechnique | Y_t, N_t] &\equiv \Pr\left[a \in \varIncident_Y | Y_t \subseteq \varIncident_Y \wedge N_t \cap \varIncident_Y = \emptyset \right] \\
    &\approx \frac{\left|\left\{ \hat{\varIncident} \in \setIncidents_{\langle Y_t,N_t\rangle} \,\middle|\, \varTechnique \in \hat{\varIncident}_Y \right\}\right|}{\left|\setIncidents_{\langle Y_t,N_t\rangle}\right|}
    \label{eq:cond_emp_prob}
\end{align}
where
\begin{align}
    \setIncidents_{\langle Y_t,N_t\rangle} = \left\{ \hat{\varIncident} \in \setIncidents \,\middle|\, Y_t \subseteq \hat{\varIncident}_Y \wedge N_t \cap \hat{\varIncident}_Y = \emptyset \right\} .
\end{align}
In other words, $\setIncidents_{\langle Y_t,N_t\rangle}$ is the set of prior incidents which ``match'' the current state of the investigation, that is, the set of prior incidents in which threat actors did use all of the techniques from $Y_t$ but did not use any of the techniques from~$N_t$.
\cref{eq:cond_emp_prob} estimates the probability $\Pr[\varTechnique \,|\, Y_t, N_t]$ as the ratio of incidents from $\setIncidents_{\langle Y_t,N_t\rangle}$ in which threat actors used technique $\varTechnique$.

\paragraph{$k$-nearest Neighbors}
The weakness of this estimation is that its accuracy depends on the cardinality of the set $\setIncidents_{\langle Y_t,N_t\rangle}$, and as the sets $Y_t$ and~$N_t$ grow with each step of the cyber-forensic investigation, the set $\setIncidents_{\langle Y_t,N_t\rangle}$ shrinks.
In fact, due to the limited number of prior incidents, the number of prior incidents that ``match'' the current state of the investigation may reach zero, which precludes us from calculating the estimates at all.
To address this issue, we extend the set $\setIncidents_{\langle Y_t,N_t\rangle}$ to include prior incidents that do not exactly ``match'' the current state of the investigation but are sufficiently similar.
We measure similarity between the current state $\langle Y_t, N_t \rangle$ and a prior incident $\hat{\varIncident} \in \setIncidents$ using a \emph{Hamming distance over the techniques} $Y_t \cup N_t$ that have already been investigated:
\begin{equation}
    d(\langle Y_t,N_t\rangle, \hat{\varIncident}) = \big| Y_t \cap \hat{\varIncident}_N \big| + \big| N_t \cap \hat{\varIncident}_Y \big| .
\end{equation}
The first term of the right-hand side is the number of techniques that were used in incident $\varIncident$ %
but not in prior incident~$\hat{\varIncident}$, while the second term is the number of techniques that were used in prior incident $\hat{\varIncident}$ but not in incident~$\varIncident$. Equivalently, we could formulate the following metric: $s(\langle Y_t,N_t\rangle, \hat{\varIncident}) = \big| Y_t \cap \hat{\varIncident}_Y \big| + \big| N_t \cap \hat{\varIncident}_N \big|$, which measures similarity by counting techniques where the current state and the prior incident are the same; in contrast, our formulation measures dissimilarity (specifically, Hamming distance) by counting techniques where the current state and the prior incident differ. These measures are practically equivalent since $d(\langle Y_t,N_t\rangle, \hat{\varIncident}) = |Y_t \cup N_t| - s(\langle Y_t,N_t\rangle, \hat{\varIncident})$%
; hence, selecting $k$ incidents with highest similarity is equivalent to selecting $k$ incidents with lowest distance (i.e., $k$-NN).%

Finally, we replace the set of ``matching'' prior incidents $\setIncidents_{\langle Y_t, N_t \rangle}$ in \cref{eq:cond_emp_prob} with the set of $k$ prior incidents that are closest with respect to metric $d$ (breaking ties arbitrarily).
Notice that this estimation of the probability $\Pr[\varTechnique \,|\, Y_t, N_t]$ is actually a $k$\emph{-nearest neighbor regression} with prior incidents $\setIncidents$ as the dataset, $d$ as the distance metric, and a binary value representing if technique $a$ was used in an incident as the output feature.
While the value of $k$ could be a constant hyper-parameter, we found that our approach performs better if $k$ varies throughout the investigation. 
Hence, we let $k = \beta_1 + \beta_2 \cdot t$, where $\beta_1$ and $\beta_2$ are hyper-parameters, which can find experimentally.

\input{pseudo}

\subsection{Monte Carlo Tree Search}
\label{subsection:Tree Search}

Since Monte Carlo tree search randomly samples sequences of actions, its coverage of the state space becomes sparser and sparser as it looks further into the future.
Therefore, when choosing between actions, it should assign more importance to the near future than the uncertain far future.
We can express this consideration by reformulating the objective as \emph{maximizing the expected discounted sum of rewards}:
\begin{align}
    \max_{\pi} \mathbb{E}  \Bigg[ \sum_{t=0}^{|\setTechniques| - 1} \gamma^t \cdot  1_{\{\varTechnique_t \in \varIncident_Y\}} \cdot %
    B_{\varTechnique_t} / C_{\varTechnique_t} %
    \Bigg|\, \varTechnique_t = \pi\left( Y_t, N_t \right)   \Bigg]
    \label{eq:problem_formulation}
\end{align}
where $\gamma \in (0, 1)$ is a \emph{temporal discount factor}: rewards obtained at step $t$ are discounted by a factor $\gamma^t$.

Note that we also replace the reward $B_{a_t}$ with the benefit to cost ratio $B_{a_t} / C_{a_t}$.
The rationale behind this is to incentivize the tree search to consider cost $C_a$ regardless of how far the investigation is from reaching a budget limit $G$; otherwise, the tree search would focus only on immediate benefit $B_a$.
Note that we found the ratio $B_{a_t} / C_{a_t}$ formulation to work best in practice (e.g., better than the more intuitive semi-MDP formulation with $\gamma^{\sum_{\tau=0}^t C_{a_\tau}}$ temporal discount and reward $B_{a_t}$).

In each step of the investigation, we run a Monte Carlo tree search (\cref{alg:treesearch}), starting from the current state~$\langle Y_t, N_t\rangle$, which outputs an action $a_t$ that is estimated to result in the maximum expected discounted sum of rewards.
To estimate the expected rewards for each action, 
the algorithm performs a number of iterations.
In each iteration, it first generates a sequence of actions and states, starting from the current state, which is a random sample of how the investigation might play out if certain actions are selected.
Then, in the back-propagation phase of the iteration, it updates its estimates of the expected rewards based on the experience of the sampled sequence.
While our algorithm follows the common principles of MCTS, it is tailored to our problem and takes advantage of the specific rules of our MDP.

\paragraph{Variables and Initialization}

Here, we describe the variables that are maintained by the algorithm throughout the iterations and how they are initialized; we will describe later how they are updated.

Variable $n[Y, N, a]$ is the number of times that the algorithm has tried action~$a$ in state $\langle Y, N \rangle$ (initialized to $0$ at the beginning of the search).
Note that this and all other initializations can be lazy, i.e., we can store these variables in a dictionary that is initially empty, and we can assign an initial value to a variable when we use it for the first time.
For ease of presentation, the pseudocode initializes all variables explicitly at the beginning.

Variable $R[Y, N, a]$ is an estimate of the expected discounted sum of rewards if we started from state~$\langle Y, N\rangle$, took action~$a$ first, and then followed an optimal policy.

Finally, variable $R[Y, N]$ is an estimate of the expected discounted sum of rewards if we started from state~$\langle Y, N\rangle$ and then followed an optimal policy.
While these variables could be initialized to $0$, we can improve the performance of the search by initializing them with an estimate:
\begin{equation*}
R[  Y, N  ] \leftarrow \hspace{-0.75em} \sum_{a \in \setTechniques \setminus (Y \cup N)} \hspace{-0.75em} \sum_{j = 0}^{|\setTechniques \setminus (Y \cup N)|-1} \hspace{-0.75em} \gamma^j \frac{ \Pr[a | Y_t, N_t] \cdot B_a / C_a }{|\setTechniques \setminus (Y \cup N)|}
\end{equation*}
Starting from state $\langle Y, N \rangle$, we can investigate techniques $\setTechniques \setminus (Y \cup N)$; however, we do not know the optimal order in which to investigate them.
Our  estimate calculates expected rewards assuming a random order: each technique $\varTechnique \in \setTechniques \setminus (Y \cup N)$ is equally likely to be investigated first (discount factor $1$), second (factor $\gamma$), third (factor $\gamma^2$), ..., and last (factor $\gamma^{|\setTechniques \setminus (Y \cup N)|-1}$).
Our estimate can be calculated very quickly, especially since the probability estimates $\Pr[a | Y_t, N_t]$ need to be calculated anyway for the back-propagation phase.
Note that for terminal states (i.e., when $\setTechniques \setminus (Y \cup N) = \emptyset$), this formula correctly calculates the expected rewards to be $0$, following the notational convention that summation over an empty set yields $0$.

\paragraph{Selection and Expansion}

In each iteration of the MCTS, we first generate a
sequence of states and actions $\langle Y_t, N_t \rangle, a_t, \langle Y_{t+1}, N_{t+1} \rangle, a_{t+1}, \langle Y_{t+2}, N_{t+2} \rangle, a_{t+2}, \ldots$, which we then use in the back-propagation phase to improve our estimates $R[Y, N]$ and $R[Y, N, a]$ for the states and actions in the sequence.
We generate the sequence starting from state $\langle Y_t, N_t \rangle$ by
alternating between selecting an action $a_i$ to take
and simulating the resulting transition to state $\langle Y_{i+1}, N_{i+1} \rangle$.
To select each action $a_i$, we use the exploration rule described in \cref{alg:ExplorationDecision},
which takes a state $\langle Y, N \rangle$ and outputs a selected action $a$,
balancing exploration and exploitation.
This rule is based on the commonly used UCT (Upper Confidence Bound 1 applied to trees) formula~\cite{kocsis2006bandit} with a myopic pruning (see below).
The first term $R[Y, N, a]$ gives preference to actions that should be selected because---according to our current estimates---they lead to high expected rewards;
while the second term gives preference to actions that have been explored less (i.e., tried fewer times) than other actions in the given state.
The exploration factor $M$ is a constant hyper-parameter that balances exploration and exploitation, which we can find experimentally.
Note that we add $1$ to the divisor to avoid division by zero since $n[Y, N, a]$ is initialized to $0$; however, this is just for ease of exposition, the addition of $1$ is negligible as we run a large number of iterations. %

After selecting an action $a_i$,
we update the number of times $n[Y_i, N_i, a_i]$ that action $a_i$ has been tried, and we simulate the random transition to state $\langle Y_{i+1}, N_{i+1} \rangle$.
However, instead of basing the transition on the actual probability $\Pr[a_i | Y_i, N_i]$, we select one of the two possible transitions with the same probability (i.e., $0.5$ probability that $a_i$ was used, and $0.5$ probability that it was not).
We use uniform probabilities for two reasons.
First, we factor in the actual probability $\Pr[a_i | Y_i, N_i]$ during back-propagation; hence, we obtain unbiased estimates regardless of the selection probabilities.
Second, uniform probabilities lead to a more balanced and thorough exploration of the transitions;
we found that with the actual probabilities, transitions that have high impact but low probability are not explored enough.

We stop generating the sequence of states and actions when we reach either a terminal state ($Y_i \cup N_i = \setTechniques$) or the depth limit ($i \geq t + D$, where $D$ is a hyper-parameter).

\paragraph{Myopic Pruning}

To improve search performance, %
we prune the search tree during selection by exploring only those actions that are most attractive in terms of the expected immediate reward.
Specifically, in state $\langle Y, N \rangle$, \cref{alg:ExplorationDecision} explores only those $F$ actions that have the highest $\Pr[\varTechnique \,|\, Y, N] \cdot B_\varTechnique / C_\varTechnique$, where $F$ is a hyper-parameter. 
While myopically focusing on actions with the highest expected immediate reward may occasionally disregard optimal actions, it typically provides better results by letting the search thoroughly explore the most promising branches of the tree.

\paragraph{Backpropagation}
During the back-propagation phase of each iteration,
we loop over the sequence of states and actions backwards,
and we update our estimates based on the experience of this sequence.
For each action $a_i$, we update the estimate $R[Y_i, N_i, a_i]$ using the following formula:
\begin{align*}
 R[ & Y_{i}, N_{i}, a_{i}  ] \leftarrow \nonumber\\
&\Pr[a_{i} | Y_{i}, N_{i}] \cdot (B_{a_{i}} / C_{a_{i}} + \gamma \cdot R[  Y_{i} \cup \{a_{i}\}, N_{i}  ])  \nonumber\\
& + (1 - \Pr[a_{i} | Y_{i}, N_{i}]) \cdot (\gamma \cdot  R[  Y_{i}, N_{i}\cup \{a_{i}\}  ]) .
\end{align*}
This formula calculates a best estimate of the expected discounted sum of rewards (relying on the $k$-NN based probability estimate $\Pr[a_i | Y_i, N_i]$):
if technique $a_i$ was used in the incident, 
we obtain immediate reward $B_{a_i} / C_{a_i}$ and then continue from state $\langle Y_{i} \cup \{a_{i}\}, N_{i}\rangle$, factoring in discount $\gamma$ as this state is one step into the future;
if technique $a_i$ was not used, a similar argument applies.
Note that each $R[Y, N]$ is either the initial estimate, which is~$0$ for terminal states, or an estimate based on prior iterations (see below).

For each state $\langle Y_i, N_i\rangle$, we update the estimate $R[Y_i, N_i]$ using the following formula:
\begin{align}
    R[ Y_{i},N_{i} ] \leftarrow  \max_{a \in \setTechniques \setminus (Y_{i} \cup N_{i} )} R[ Y_{i}, N_{i}, a ]
\end{align}
This formula calculates a best estimate since in each state~$\langle Y_i, N_i\rangle$, we should take the optimal action $a$ that maximizes the expected discounted sum of rewards (based on our best estimates $R[ Y_{i}, N_{i}, a]$).

%% file: pseudo.tex
\begin{algorithm}[t]
\SetKwInput{KwInput}{Input}               
\SetKwInput{KwOutput}{Output}  
\SetAlgoLined
\SetKwProg{Fn}{Function}{:}{}
  \Fn{{\textit{ExplorationDecision}$(Y, N, M,F)$}}{
    $\textit{MP} \leftarrow \argmax_{\substack{\mathcal{A}' \subseteq \mathcal{A} \setminus {Y \cup N},\\ |\mathcal{A'}| = F}} \sum_{a \in \mathcal{A}'} \Pr[\varTechnique \,|\, Y, N] \cdot B_\varTechnique / C_\varTechnique $\\
      $n[ Y, N ] \leftarrow \sum_{a \in \mathcal{A} \setminus (Y \cup N)} n[ Y, N, a ]$\\
 \Return $\argmax\limits_{a \in \textit{MP}} R[ Y,N,a ] + M \sqrt{\frac{ \ln n[  Y, N ]  }{ n[  Y, N, a  ] + 1 } }$}
~~~\caption{Exploration Decision}
\label{alg:ExplorationDecision}
\end{algorithm}

\begin{algorithm}[t]
\SetKwInput{KwInput}{Input}{\textbf{Input}: {state} $\langle Y_t, N_t \rangle$, {constants} $\mathcal{A}, \boldsymbol{B}, \boldsymbol{C}, \gamma, K, D, M, F$
}\\                
\SetKwInput{KwOutput}{Output}{\textbf{Output:} action $a_t$}\\
\textbf{Initialization}: \\
$\forall  \langle Y, N, a \rangle: n[  Y, N, a  ] \leftarrow 0$ \\
$\forall  \langle Y, N, a \rangle: R[  Y, N, a  ] \leftarrow 0$ \\
$\forall  \langle Y, N \rangle: R[  Y, N  ] \leftarrow$ \\  
$~~~~~~~~~~~~~~\sum_{a \in \setTechniques \setminus (Y \cup N)} \sum_{j = 0}^{|\setTechniques \setminus (Y \cup N)|-1} \gamma^j \frac{ \Pr[a | Y_t, N_t] \cdot B_a / C_a }{|\setTechniques \setminus (Y \cup N)|}$ \\ 

\For{$K$ times}{
$i \leftarrow t$ \\
\While{$Y_{i} \cup N_{i} \neq \setTechniques \textnormal{\textbf{ and }} i < t + D$}{
$a_{i} \leftarrow \textit{ExplorationDecision}(Y_{i}, N_{i}, M, F)$ \\
$n[ Y_{i}, N_{i}, a_{i}  ] \leftarrow n[  Y_{i}, N_{i}, a_{i}  ] + 1$ \\
\eIf{$random(0,1) < 0.5$}{
  $Y_{i+1} \leftarrow Y_{i} \cup \{a_{i}\}$ \\  
  $N_{i+1} \leftarrow N_{i}$
}{
  $Y_{i+1} \leftarrow Y_{i}$ \\ 
  $N_{i+1} \leftarrow N_{i}$ $\cup$ $\{a_{i}\}$
}
$i \leftarrow i + 1$
}
$i \leftarrow i - 1$ \\ 
\While{$i\geq t$}{
$R[  Y_{i},N_{i},a_{i}  ] \leftarrow$ \\ 
$~~~~~ \Pr[a_{i} | Y_{i}, N_{i}] \cdot (B_{a_{i}} / C_{a_{i}} + \gamma \cdot R[  Y_{i} \cup \{a_{i}\}, N_{i}  ])$ \\
$~~~~~ + (1 - \Pr[a_{i} | Y_{i}, N_{i}]) \cdot \gamma \cdot  R[ Y_{i}, N_{i}\cup \{a_{i}\} ]$ \\
{$R[ Y_{i},N_{i} ] \leftarrow  \max_{a \in \setTechniques \setminus (Y_{i} \cup N_{i} )} R[ Y_{i}, N_{i}, a ]$} \\
$i \leftarrow i - 1$ 
}
}
$a_t \leftarrow \argmax_{a \in \setTechniques \setminus (Y_t \cup N_t) } R[ Y_t, N_t, a ]$ \\
\caption{MCTS for Forensic Decision Support}
\label{alg:treesearch}
\end{algorithm}

%% file: results.tex
\input{plots/benefit_cost_budget_45}
\input{plots/benefit_cost_budget_65}
\section{Numerical Evaluation}
\label{sec:results}
We evaluate our proposed approach numerically on public datasets of real-world cyber incidents.
For each cyber incident, we simulate how our approach would have prioritized the investigation, and plot the benefit obtained through discovery as a function of the effort spent.
We compare our approach to two baselines, DISCLOSE and a na\"ive policy.
Our implementation and datasets are publicly available.\footnote{\url{https://github.com/SoodehAtefi/DecisionSupport-AAAI-23}}

\subsection{Experimental Setup}

\paragraph{Dataset}
We use the MITRE ATT\&CK Enterprise repository \cite{barnum2012standardizing},
which is a public repository of adversarial tactics, techniques, \& procedures, referencing real-world cyber incidents in which some of these techniques were used.
Since its original publication, both the ATT\&CK framework and its CTI dataset have been regularly updated. 
We use three versions of the repository in our evaluation: v6.3 (2019), which is the version that Nisioti et al.~\shortcite{nisioti2021data} used to evaluate DISCLOSE; v10.1 (2021); and v11.3 (2022), which is the latest version at the time of writing. %
Evaluating our approach on multiple versions demonstrates that it can be applied to newer versions without any changes (other than standard hyper-parameter optimization). 

For a fair comparison, we use the same 31 techniques~$\setTechniques$ and the same benefit $B$ and cost $C$ values as Nisioti et al.~\shortcite{nisioti2021data}. Since the categorization of techniques changed slightly between versions, we had to map some techniques to equivalent ones for later versions.  This leaves us with 29 techniques for versions v10.1 and v11.3.
The benefit and cost values are based on the Common Vulnerability Scoring System and interviews with cyber-forensic experts.

For each technique, we collected cyber incidents in which the technique was used via the \texttt{external\_references} field (replicating Nisioti et al.~\shortcite{nisioti2021data}). 
Our v6.3, v10.1, and v11.3 datasets contain 331, 670, and 716 cyber incidents, respectively.
Version v11.3 includes 425 incidents that are new compared to v6.3, and 73 incidents that are new compared to v10.1. 
In these datasets, every incident used at least 2 techniques. In dataset v11.3, incidents used 4.1 techniques on average (min. 2; max. 17). One example of an incident that used many techniques is the Frankenstein campaign~\cite{biasini2019it}, which employed Phishing, OS Credential Dumping, Exfiltration Over C2 Channel, and other techniques.

\paragraph{Baselines}
We compare our approach to two baselines, DISCLOSE (implemented as specified in Nisioti et al.~\shortcite{nisioti2021data})  and a na\"ive baseline, which we call the \emph{static policy}.
At every step, the static policy selects the technique (from the set of techniques that have not been investigated yet) that is most frequent across all prior incidents, instead of considering conditional probabilities.
This na\"ive baseline represents investigation without decision support (i.e., decisions that ignore the state).

\paragraph{Simulation Setup and Metrics}
Since the datasets are relatively small, we use a leave-one-out cross-validation: 
when evaluating a policy on an incident, we treat all other incidents in our dataset as prior incidents $\setIncidents$.
We always simulate the investigation starting with a single randomly chosen technique from $\varIncident_Y$ as the singleton set $Y_0$ (and letting $N_0 = \emptyset$).
Same as Nisioti et al.~\shortcite{nisioti2021data}, we terminate an investigation when the \emph{cumulative effort cost} (i.e., $\sum_{\varTechnique \in Y_t \cup N_t} C_\varTechnique$) reaches 45 or 65 to assess which techniques could have been promptly discovered by the approach. We also conducted experiments with two more budget limits (70 and 100) and without a budget limit to provide a more comprehensive comparison (we include results in {\ifExtendedVersion \cref{sec:different-budgets}\else the online appendix~\cite{atefi2022principled}\fi}). %
Further, to quantify promptness, we measure the \emph{area under the benefit-effort curve} (AUCBE):
we plot the  discovery of benefits obtained 
as a function of the cumulative effort cost for three different approaches (e.g., see \cref{fig:BenefitCost45v6}). Higher AUCBE values are~better. Solid lines are averages over all incidents, dotted lines are 25\% quantiles, and dashed lines are 75\% quantiles.
\Aron{quality of writing?}%

\paragraph{Hyperparameter Optimization}
While the baseline does not have any hyper-parameters, our proposed approach has a number of them.
First, we optimized the hyper-parameters for the $k$-NN probability estimation ($\beta_1, \beta_2$) using a grid search, maximizing the average AUCBE over all incidents.
During this search, we restricted the MCTS to one-action depth ($D = 1$, in which case the other parameters of the MCTS are irrelevant), providing us with good hyper-parameters for probability estimation.
Then, we optimized the hyper-parameters for MCTS using Hyperopt Python library, again maximizing AUCBE.
Note that we optimized the hyper-parameters for datasets separately.
We provide results from the hyper-parameter search in \ifExtendedVersion \cref{sec:parameter-search}\else the online appendix~\cite{atefi2022principled}\fi.

\subsection{Numerical Results}

\paragraph{Running Time}
For a single decision, the MCTS algorithm takes less than 7 seconds on average using a single core of a 2.4GHz Intel Core i9 CPU, and less than a second using multiple cores. %
Compared to the amount of time that forensic analysts need to investigate the selected technique (at the very least minutes), these running times are negligible.

\paragraph{Benefits of Prioritization}

\cref{fig:BenefitCost45v6,fig:BenefitCost65v6} show cumulative benefits obtained during the cyber-forensic investigations (i.e., $\sum_{\varTechnique \in Y_t} B_\varTechnique$) as functions of the cumulative effort costs (i.e., $\sum_{\varTechnique \in Y_t \cup N_t} C_\varTechnique$) using MCTS, DISCLOSE, and static approach on the v6.3 dataset. We conducted the experiments with budgets 45 and 65. We provide results for other versions of the dataset and different budgets in \ifExtendedVersion \cref{sec:different-budgets}\else the online appendix~\cite{atefi2022principled}\fi.
Each curve is based on all incidents in the dataset.
For the v6.3 dataset, our approach outperforms the baselines in both of the scenarios (45, and 65). The average AUCBE of MCTS, DISCLOSE, and static policy for budget limit 45 are 3,503, 3,411, and 3,175 respectively. The average AUCBE of our approach, DISCLOSE, and static policy for budget limit 65 are 6,288, 6,108, and 5,826, respectively.

For the v11.3 dataset, our approach outperforms the baselines in both of the scenarios as well. The average AUCBE of MCTS, DISCLOSE, and static policy for budget limit 45 are 4,061, 3,982, and 3,865, and for budget limit 65 are 7,072, 6,975, and 6,816 respectively.  
This demonstrates that while our principled approach provides superior prioritization, the decision-support problem is very challenging due to the inherent uncertainty of cyber incidents.

%% file: plots/benefit_cost_budget_45.tex
\pgfplotstableread[col sep=comma,]{data/results/result_benefit_v6_31.csv}\ResultsFirstVersion

\begin{figure}[t]
\begin{tikzpicture}
	\begin{axis}[font=\small,legend style={font=\tiny},
	    xmin=0,xmax=45,ymin=0,ymax=180,
		xlabel=Cumulative Effort Cost,
        ylabel=Cumulative Benefit Obtained,
            width=\linewidth,height = 5.8cm, 
        no markers,
        legend pos=south east,
		legend style={at={(0.15,1)},anchor=north}
		]

		      \addplot [semithick, color = red ]table [x=Budget, y=DISCLOSE_Benefit_45] {\ResultsFirstVersion};
      \addlegendentry{DISCLOSE}
    
		      \addplot [semithick, color = green ]table [x=Budget, y=Static_Benefit_45] {\ResultsFirstVersion};
      \addlegendentry{Static}
		 \addplot [semithick,color = blue ]table [x=Budget, y=MCTS_Benefit_45] {\ResultsFirstVersion};
      \addlegendentry{MCTS}

\addplot [semithick,dotted, color = red ]table [x=Budget, y=DISCLOSE_45_0.25] {\ResultsFirstVersion};
		      \addplot [semithick, dotted,color = green ]table [x=Budget, y=Static_45_0.25] {\ResultsFirstVersion};
	      \addplot [semithick,dotted,color = blue ]table [x=Budget, y=MCTS_45_0.25] {\ResultsFirstVersion};

      \addplot [semithick, dashed,color = red ]table [x=Budget, y=DISCLOSE_45_0.75] {\ResultsFirstVersion};
		      \addplot [semithick,dashed, color = green ]table [x=Budget, y=Static_45_0.75] {\ResultsFirstVersion};
	      \addplot [semithick,dashed,color = blue ]table [x=Budget, y=MCTS_45_0.75] {\ResultsFirstVersion};
	\end{axis}
\end{tikzpicture}

\centering
		\caption{Cumulative benefit obtained  as a function of cumulative effort cost ({up to budget 45}) on dataset {v6.3}.}
	\label{fig:BenefitCost45v6}
\end{figure}
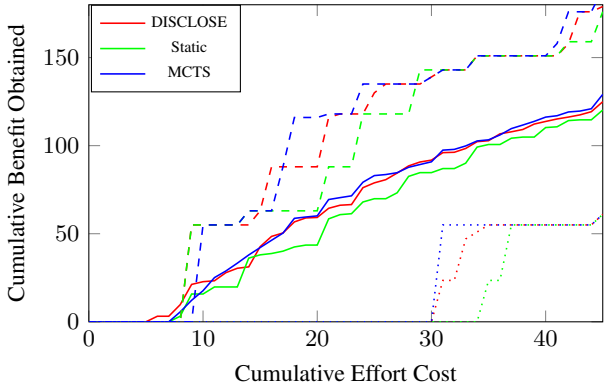

%% file: plots/benefit_cost_budget_65.tex
\pgfplotstableread[col sep=comma,]{data/results/result_benefit_v6_31.csv}\ResultsFirstVersion

\begin{figure}[t]
\begin{tikzpicture}
	\begin{axis}[font=\small,legend style={font=\tiny},
	    xmin=0,xmax=65,ymin=0,ymax=180,
		xlabel=Cumulative Effort Cost,
        ylabel=Cumulative Benefit Obtained,
            width=\linewidth,height = 5.8cm, 
        no markers,
        legend pos=south east,
		legend style={at={(0.15,1)},anchor=north}
		]

		      \addplot [semithick, color = red ]table [x=Budget, y=DISCLOSE_Benefit_65] {\ResultsFirstVersion};
      \addlegendentry{DISCLOSE}
    
		      \addplot [semithick,color = green ]table [x=Budget, y=Static_Benefit_65] {\ResultsFirstVersion};
      \addlegendentry{Static}
	      \addplot [semithick,color = blue  ]table [x=Budget, y=MCTS_Benefit_65] {\ResultsFirstVersion};
      \addlegendentry{MCTS}

\addplot [semithick,dotted, color = red ]table [x=Budget, y=DISCLOSE_65_0.25] {\ResultsFirstVersion};
		      \addplot [semithick, dotted,color = green ]table [x=Budget, y=Static_65_0.25] {\ResultsFirstVersion};
	      \addplot [semithick,dotted,color = blue ]table [x=Budget, y=MCTS_65_0.25] {\ResultsFirstVersion};

      \addplot [semithick, dashed,color = red ]table [x=Budget, y=DISCLOSE_65_0.75] {\ResultsFirstVersion};
		      \addplot [semithick,dashed, color = green ]table [x=Budget, y=Static_65_0.75] {\ResultsFirstVersion};
	      \addplot [semithick,dashed,color = blue ]table [x=Budget, y=MCTS_65_0.75] {\ResultsFirstVersion};
	\end{axis}
\end{tikzpicture}
		\caption{Cumulative benefit obtained as a function of cumulative effort cost ({up to budget 65}) on dataset {v6.3}.}
	\label{fig:BenefitCost65v6}
\end{figure}
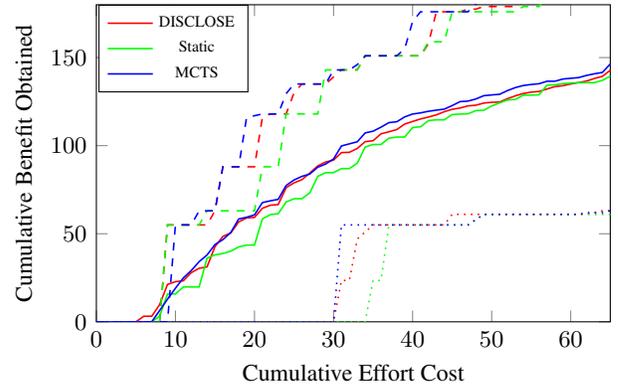

%% file: related.tex
\section{Related Work}
\label{sec:related}

Several prior research efforts focused on enhancing the efficacy of forensic investigations by decreasing the time or resources required for an investigation without impacting its quality and objectivity~\cite{hossain2018dependence,hossain2017sleuth,hossain2020combating,hassan2020omegalog,satvat2021extractor}. %
Our approach is primarily motivated by DISCLOSE \cite{nisioti2021data}, a data-driven decision-support framework, which creates \TTP space using the MITRE ATT\&CK dataset, computes conditional probabilistic relations and proximity values between \TTPs, and proposes actions based on these relations. 
DISCLOSE can assist an expert in each step of the investigation by considering benefits, costs, and available budget. 
We extend this work by introducing a formal model of the cyber-forensic investigation process, modeling it as a Markov decision process.
Further, we propose a computational approach which,  at each step of the investigation, considers all the techniques that have been investigated, instead of considering only the very last technique---as DISCLOSE~does.

Horsman et al.~\shortcite{horsman2014case} present CBR-FT, a case driven reasoning based technique for device triage. 
Similar to our approach, CBR-FT uses a knowledge base of past cases to calculate probable subsequent actions based on system file paths, which is then used to offer prediction of triage process of the current case under investigation.
Unlike the system file paths, we use TTPs which allows for more flexible analysis and reasoning of adversarial behavior.
Attack graphs were included in forensic investigation by Lui et al.~\shortcite{liu2012using} to guide the decision process. 
The notion of anti-forensic steps, parallel to other adversarial actions, were included in the attack graphs to represent that the adversary may hide relevant forensic evidence to dissuade the defender. 
Likewise, Nassif and Hruschka \shortcite{da2012document} propose the use of clustering techniques to support forensic investigation. 
Using a similar approach, Barrere et al.~\shortcite{barrere2017tracking} proposed an algorithm using condensed attack graphs that transformed the structure of the original attack graph allowing for more effective exploration.
Algorithms are also proposed to support forensic investigation of online social networks \cite{arshad2020semi} and for unsupervised prediction for proactive forensic investigation of insider threats \cite{wei2021insider}.
Similar to our work, these algorithms aim to increase the efficacy of forensic investigation by decreasing the investigation time; however, they approach the problem differently by decreasing the time required for crucial tasks.   

Quick and Choo \shortcite{quick2014impacts} highlight the effects of large amounts of digital-forensic data on modern forensic investigations.  
Saeed et al.~\shortcite{saeed2020survey} and Nisioti et al.~\shortcite{nisioti2021game} present game-theoretic approaches used to model interactions between an investigator and an adversary using anti-forensic method. 
By finding the Nash equilibrium of the game, the authors find the optimal trade-off strategy for each player. 
Even though our work does not explicitly model anti-forensic techniques and their analysis, our approach is developed considering the potential existence of anti-forensic techniques for an incident under investigation.

Defenders face a similar prioritization problem with sensitive intrusion-detection systems, which may raise a large number of false  alarms~\cite{tong2020finding,yan2019database,yan2018get,laszka2017game}.

%% file: concl.tex
\section{Discussion and Conclusions}
\label{sec:concl}

To address the limitations of DISCLOSE, we introduced a principled approach by modeling cyber-forensic investigation as Markov decision processes and proposing an MCTS and $k$-NN regression based computational approach.
A key advantage of our proposed approach is that it works directly with the data (i.e., at every step and every iteration, probabilities are estimated based on the dataset of all prior incidents), instead of training a parametric machine-learning model.
By relying on non-parametric machine-learning models, we aim to get as much ``mileage'' out of the data as possible, which is both enabled and necessitated by the limited amount of public data about cyber incidents.
Another key advantage of our proposed approach is that the tree search approximates best estimates---and hence optimal decisions---based on the dataset (as the number of iterations increases).

While these advantages enabled our approach to outperform baselines, including DISCLOSE, we found the prioritization problem to be very challenging.
The primary reason for this is the inherent difficulty of predicting how threat actors work.
In fact, DISCLOSE itself is not too far from the na\"ive baseline policy.

%% file: appendix.tex
\input{plots/benefit_cost_budget_45_v10}
\input{plots/benefit_cost_budget_45_v11}
\input{plots/benefit_cost_budget_65_v10}
\input{plots/benefit_cost_budget_65_v11}
\input{plots/benefit_cost_budget_70_v6}
\input{plots/benefit_cost_budget_70_v10}
\input{plots/benefit_cost_budget_70_v11}
\input{plots/benefit_cost_budget_100_v6}
\input{plots/benefit_cost_budget_100_v10}
\input{plots/benefit_cost_budget_100_v11}
\input{plots/benefit_cost_without_budget_v6}
\input{plots/benefit_cost_without_budget_v10}
\input{plots/benefit_cost_without_budget_v11}

\section{Appendix}
\label{sec:appendix}
\subsection{Evaluation with Different Budgets}
\label{sec:different-budgets}
\cref{fig:BenefitCost45v10,fig:BenefitCost65v10,fig:BenefitCost70v10,fig:BenefitCost100v10,fig:BenefitCostwithoutbudgetv10,fig:BenefitCost45v11,fig:BenefitCost65v11,fig:BenefitCost70v11,fig:BenefitCost100v11,fig:BenefitCostwithoutbudgetv11,fig:BenefitCost70v6,fig:BenefitCost100v6,fig:BenefitCostwithoutbudgetv6} 
show cumulative benefits obtained during the cyber-forensic investigations (i.e., $\sum_{\varTechnique \in Y_t} B_\varTechnique$) as functions of the cumulative effort costs (i.e., $\sum_{\varTechnique \in Y_t \cup N_t} C_\varTechnique$) using MCTS, DISCLOSE, and static approach on the v6.3, v10.1, and v11.3 datasets with budget limits of 45 and 65 (except v6.3), 70, 100, and without budget limit. We provided figures for the v6.3 dataset with budget limits 45 and 65  in the main text (\cref{sec:results}). Solid lines are averages over all incidents, dotted lines are 25\% quantiles, and dashed lines are 75\% quantiles.

\subsection{Hyper-Parameter Search}
\label{sec:parameter-search}

\begin{figure}[ht!]
  \includegraphics[width=\linewidth]{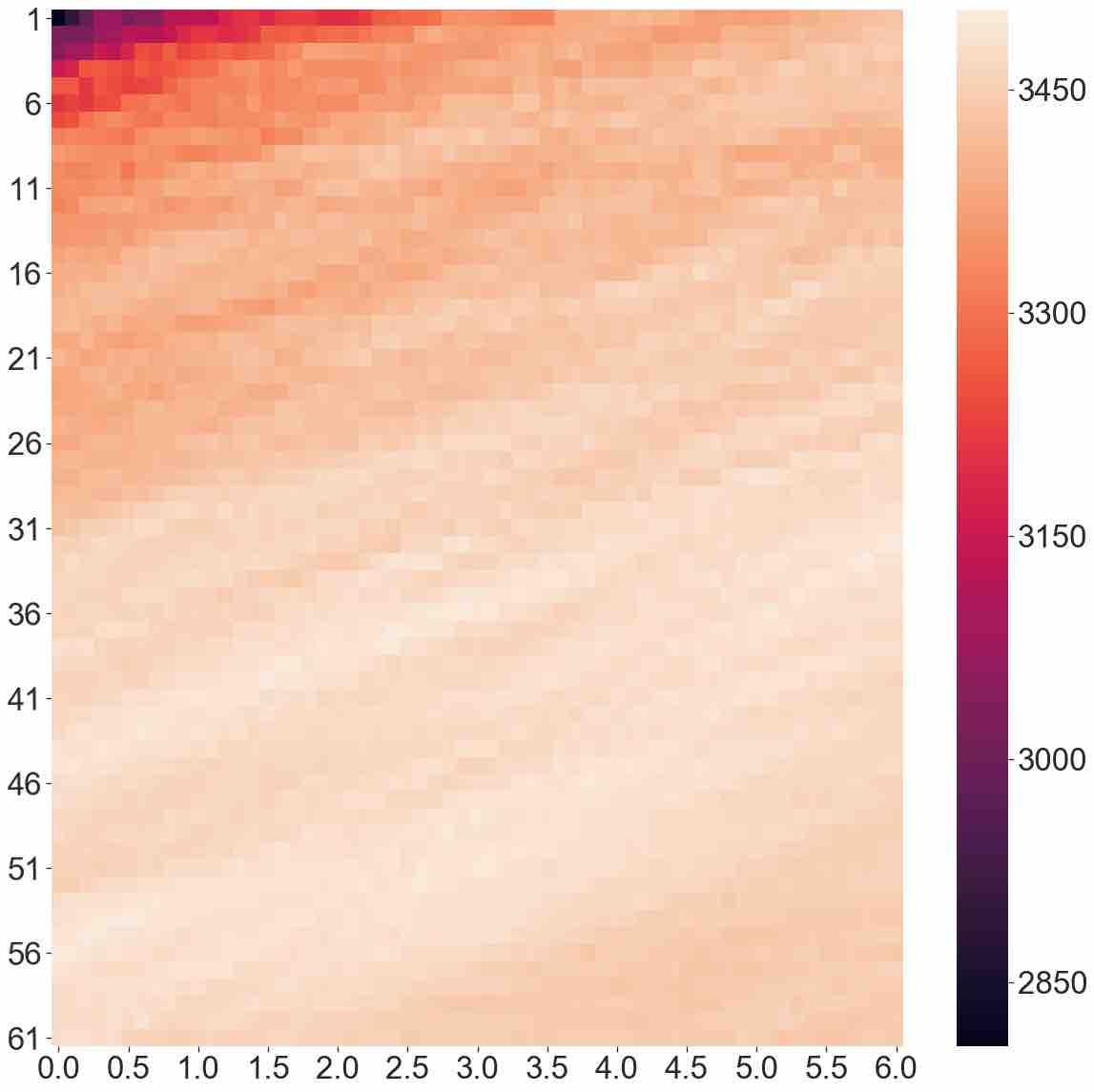}
  \caption{Benefit percentage attained AUCBE (optimal values for $\beta_1$ and $\beta_2$ are 40 and 1.5 respectively) \textbf{up to budget 45} on \textbf{v6.3}}
  \label{fig:v645}
\end{figure}

\begin{figure}[ht!]
  \includegraphics[width=\linewidth]{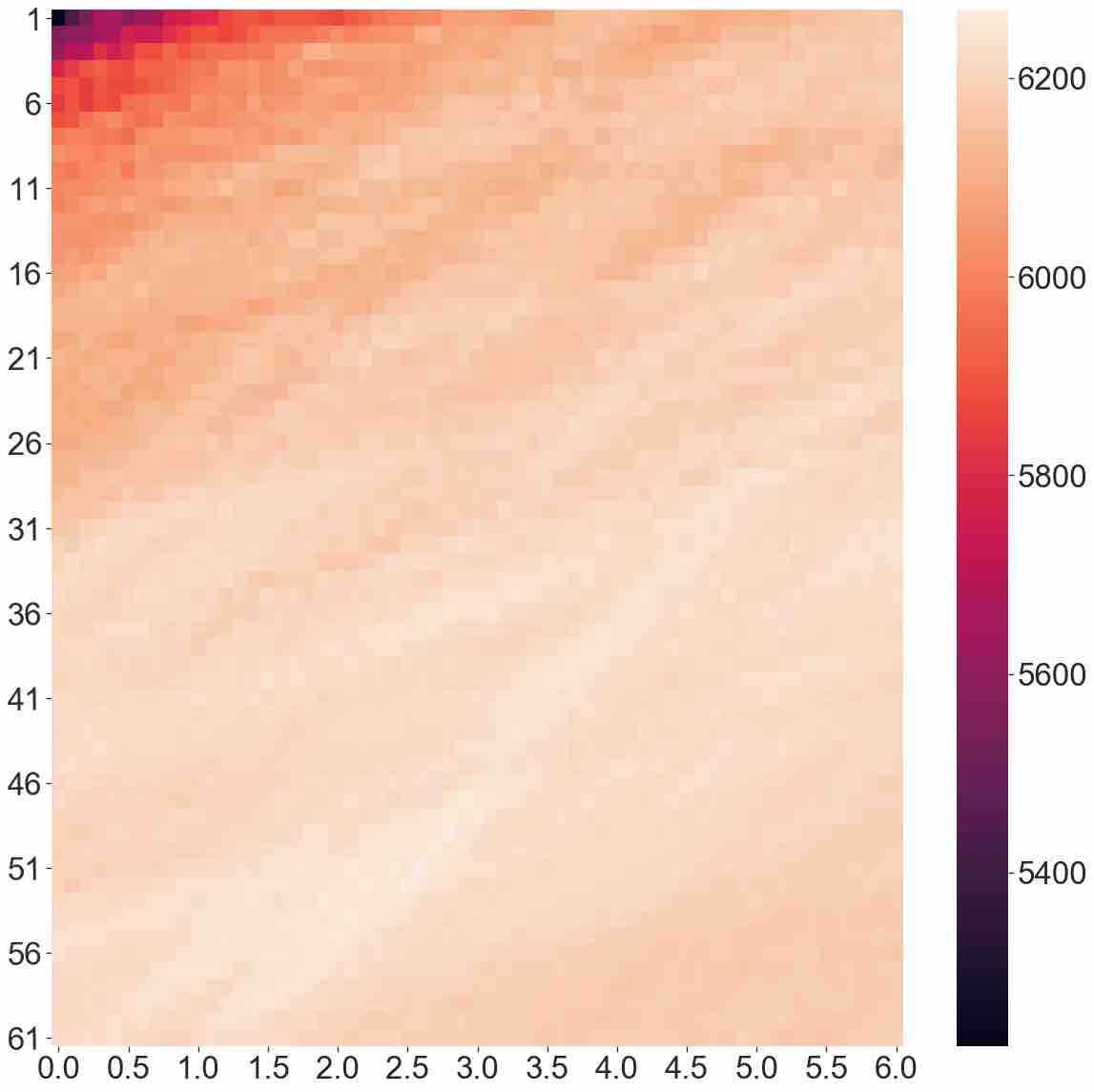}
  \caption{Benefit percentage attained AUCBE (optimal values for $\beta_1$ and $\beta_2$ are 51 and 2.6 respectively) \textbf{up to budget 65} on \textbf{v6.3}}
  \label{fig:v665}
\end{figure}

\begin{figure}[ht!]
  \includegraphics[width=\linewidth]{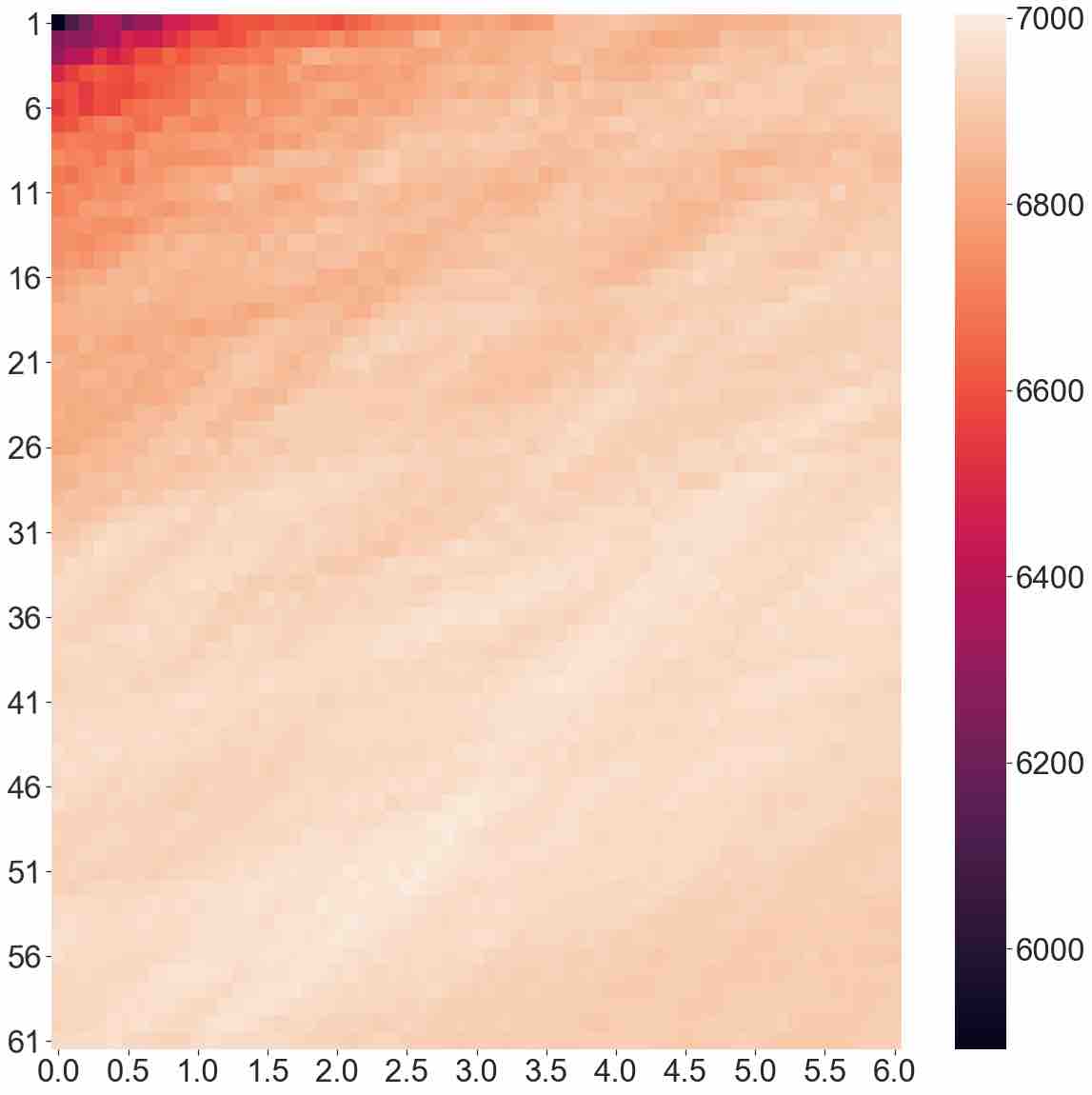}
  \caption{Benefit percentage attained AUCBE (optimal values for $\beta_1$ and $\beta_2$ are 51 and 2.6 respectively) \textbf{up to budget 70} on \textbf{v6.3}}
  \label{fig:v670}
\end{figure}

\begin{figure}[ht!]
  \includegraphics[width=\linewidth]{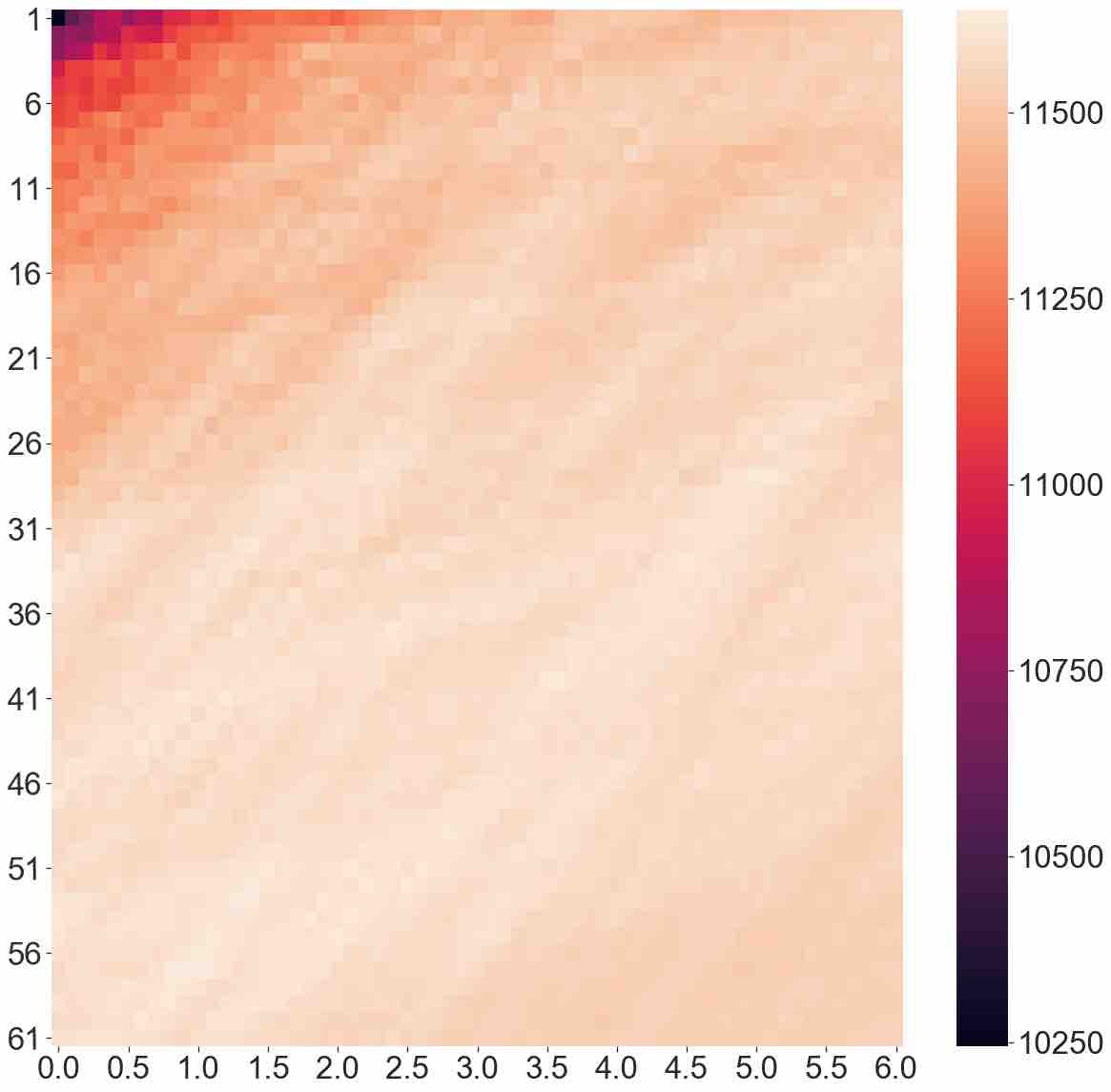}
  \caption{Benefit percentage attained AUCBE (optimal values for $\beta_1$ and $\beta_2$ are 57 and 0.9 respectively) \textbf{up to budget 100} on \textbf{v6.3}}
  \label{fig:v6100}
\end{figure}

\begin{figure}[ht!]
  \includegraphics[width=\linewidth]{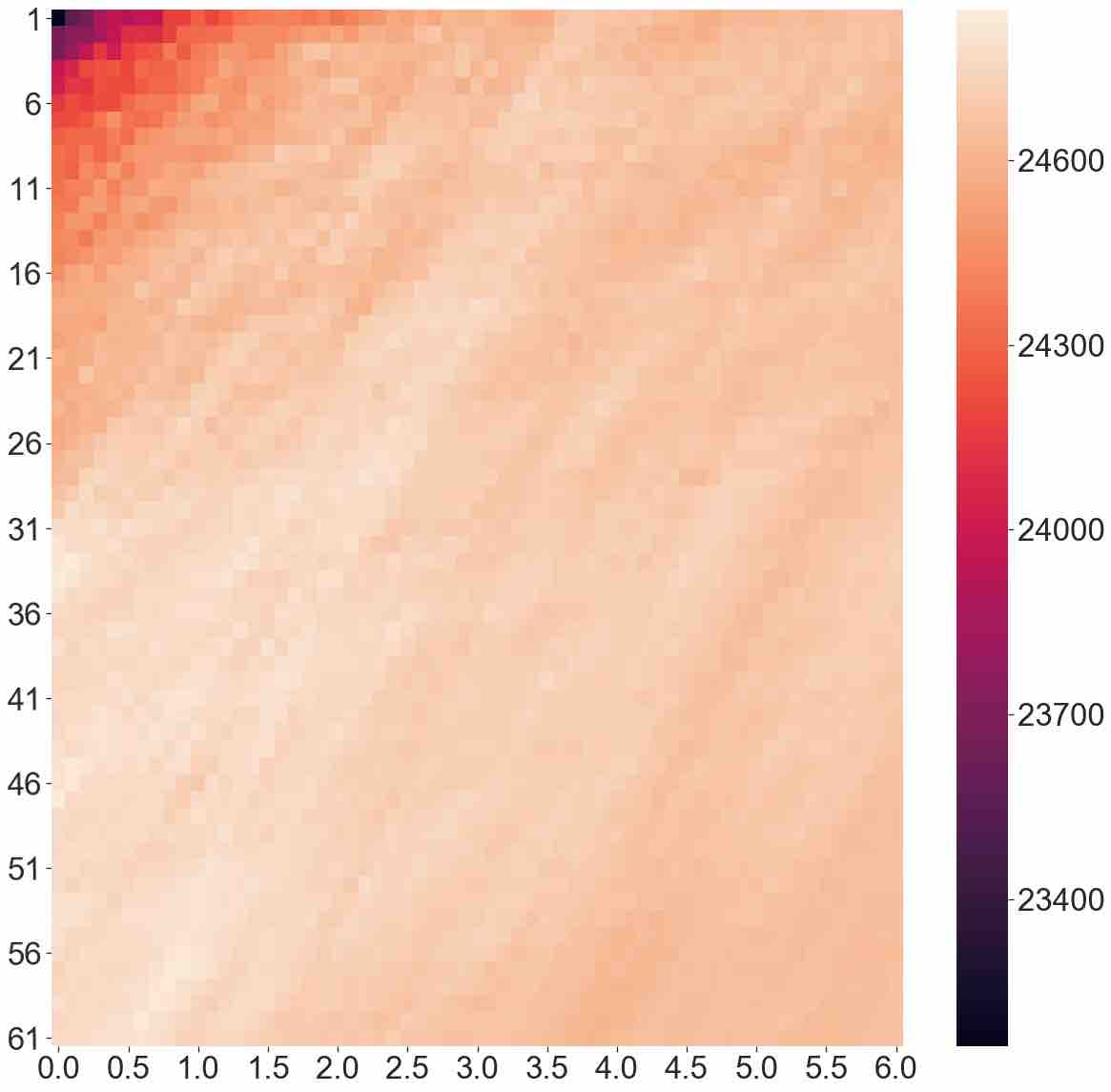}
  \caption{Benefit percentage attained AUCBE (optimal values for $\beta_1$ and $\beta_2$ are 47 and 0 respectively) \textbf{without budget limitation} on \textbf{v6.3}}
  \label{fig:v6withoutb}
\end{figure}

\begin{figure}[ht!]
  \includegraphics[width=\linewidth]{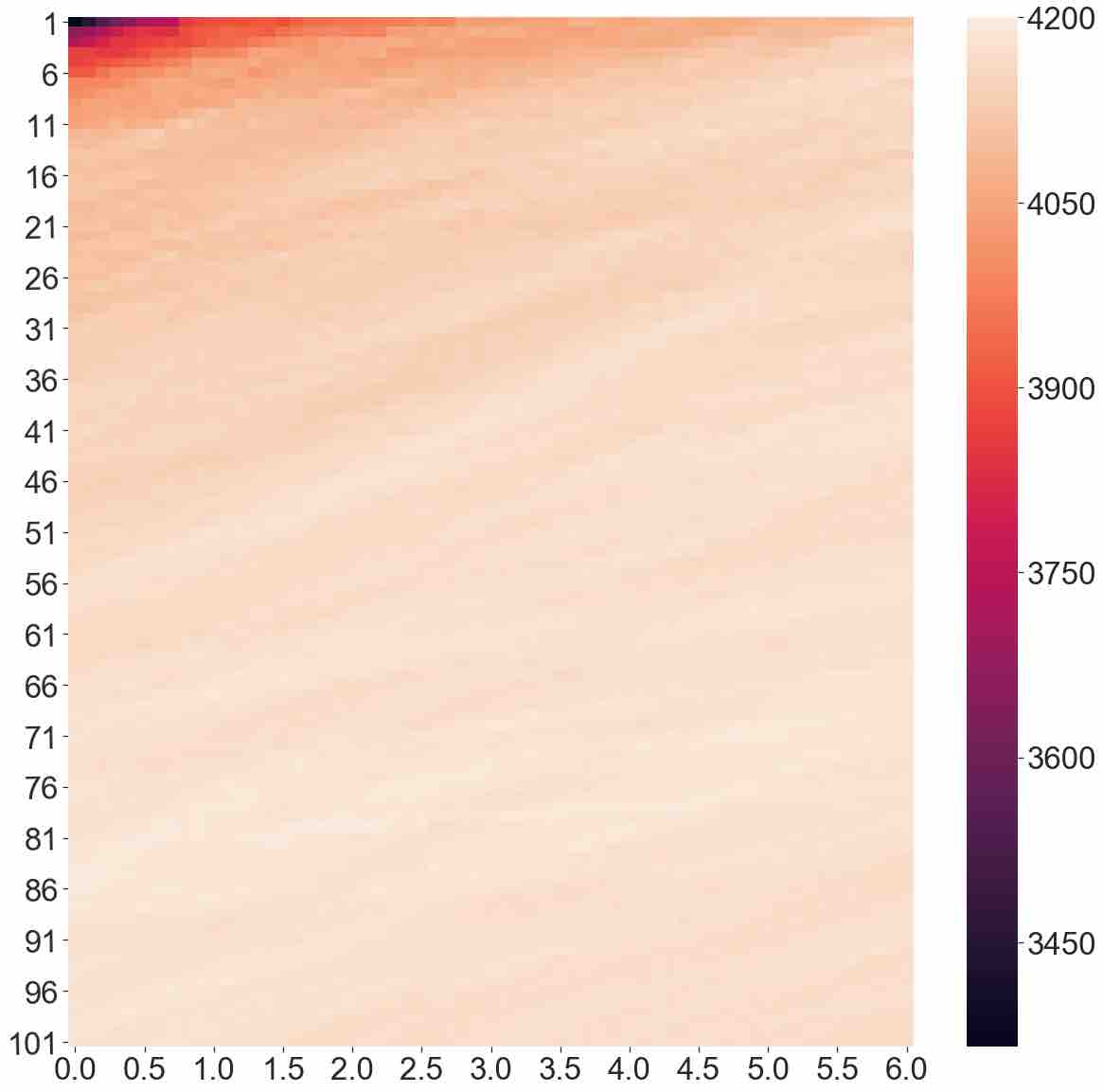}
  \caption{Benefit percentage attained AUCBE (optimal values for $\beta_1$ and $\beta_2$ are 81 and 1.4 respectively) \textbf{up to budget 45} on \textbf{v10.1}}
  \label{fig:v1045}
\end{figure}

\begin{figure}[ht!]
  \includegraphics[width=\linewidth]{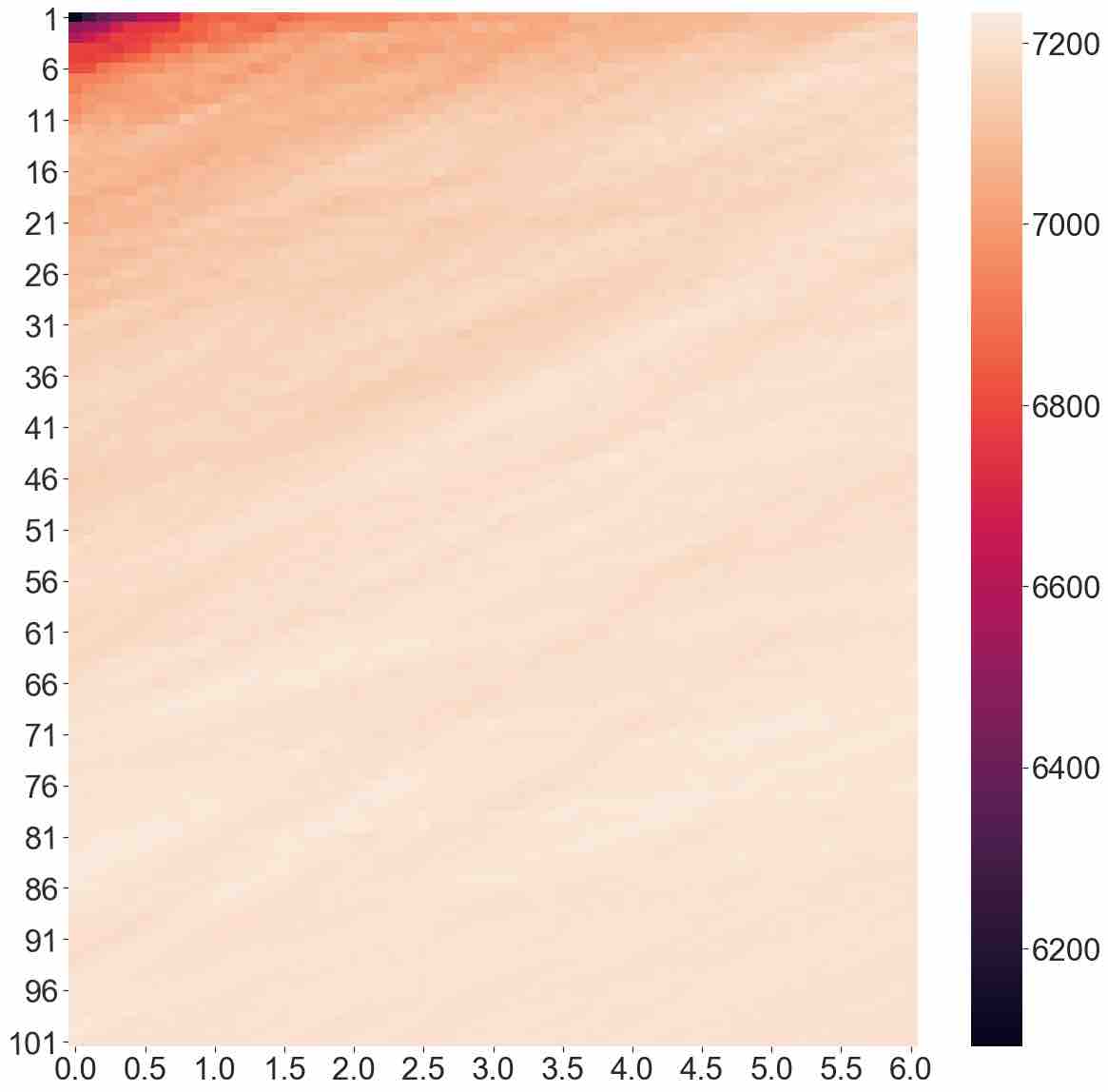}
  \caption{Benefit percentage attained AUCBE (optimal values for $\beta_1$ and $\beta_2$ are 80 and 1.8 respectively) \textbf{up to budget 65} on \textbf{v10.1}}
  \label{fig:v1065}
\end{figure}

\begin{figure}[ht!]
  \includegraphics[width=\linewidth]{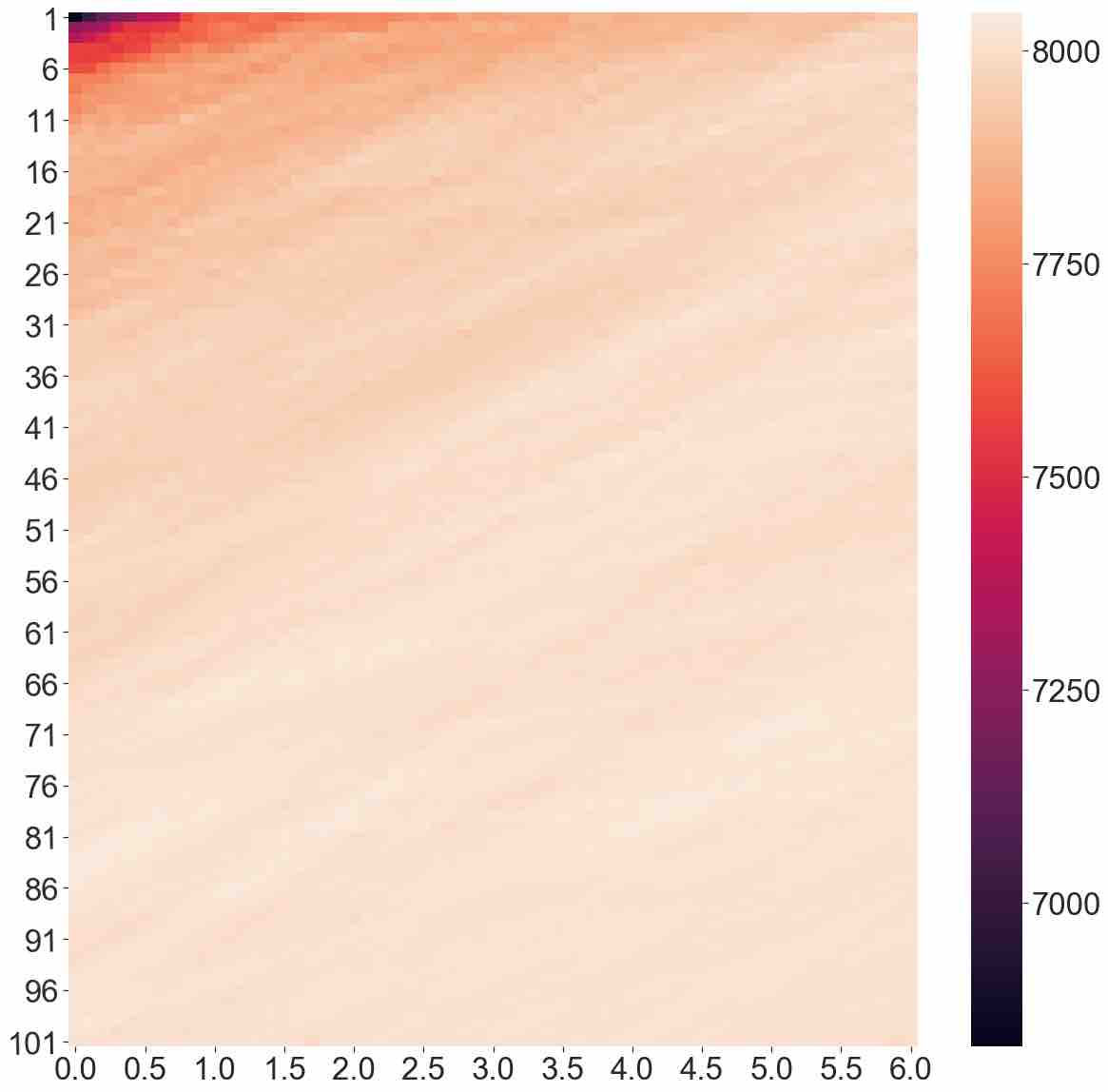}
  \caption{Benefit percentage attained AUCBE (optimal values for $\beta_1$ and $\beta_2$ are 84 and 0 respectively) \textbf{up to budget 70} on \textbf{v10.1}}
  \label{fig:v1070}
\end{figure}

\begin{figure}[ht!]
  \includegraphics[width=\linewidth]{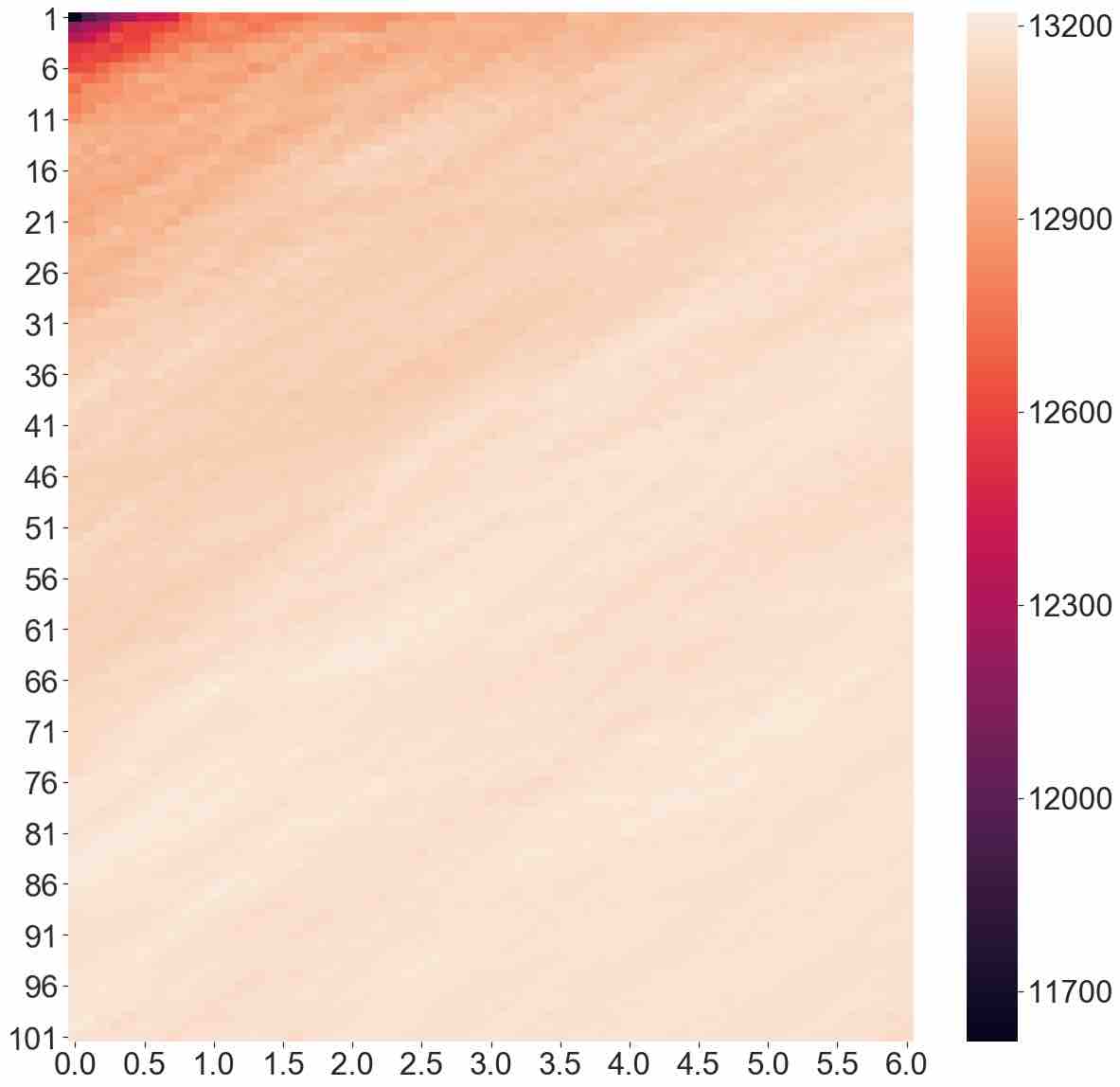}
  \caption{Benefit percentage attained AUCBE (optimal values for $\beta_1$ and $\beta_2$ are 82 and 0.2 respectively) \textbf{up to budget 100} on \textbf{v10.1}}
  \label{fig:v10100}
\end{figure}

\begin{figure}[ht!]
  \includegraphics[width=\linewidth]{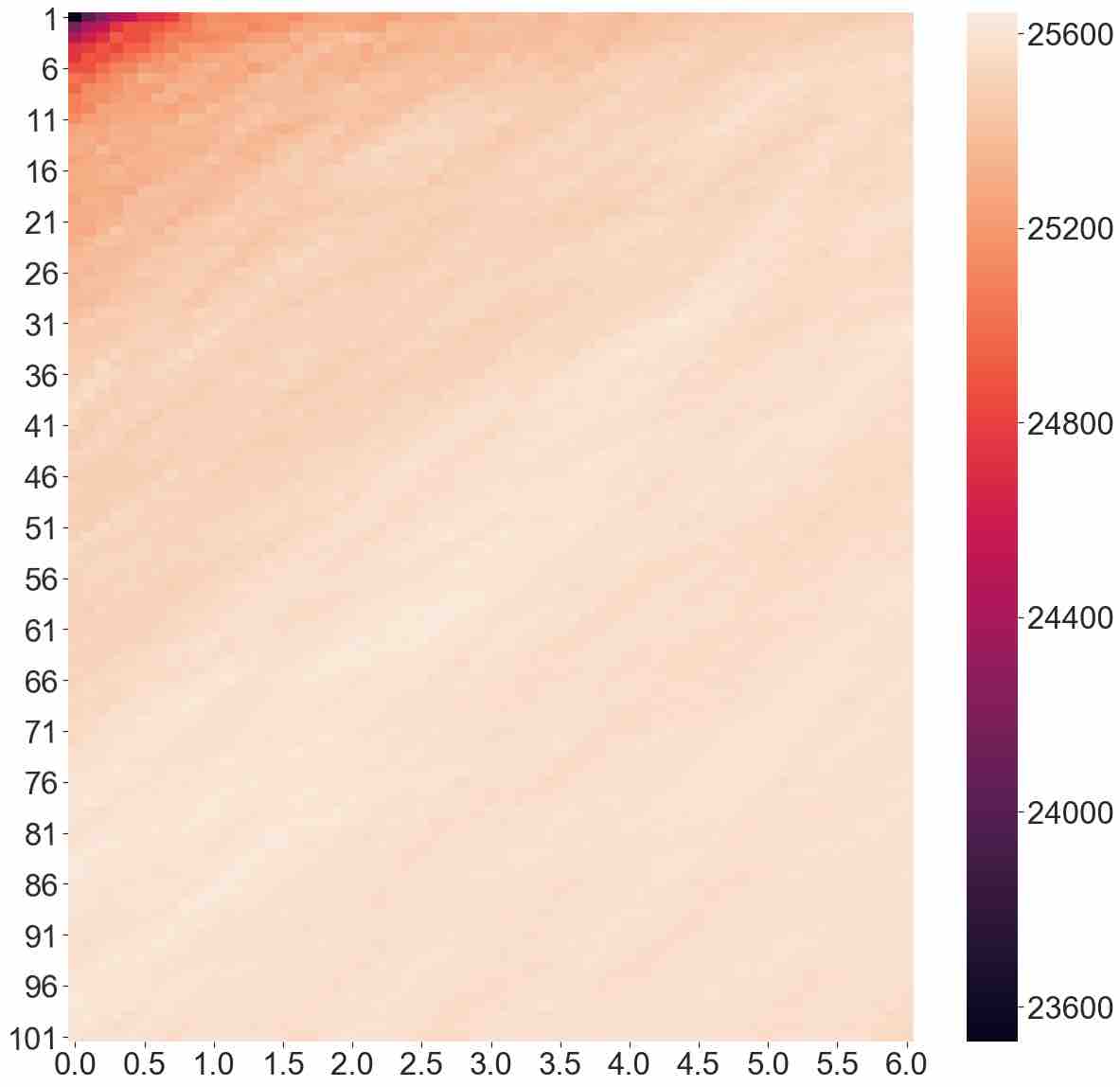}
  \caption{Benefit percentage attained AUCBE (optimal values for $\beta_1$ and $\beta_2$ are 87 and 1 respectively) \textbf{without budget limitation} on \textbf{10.1}}
  \label{fig:v10withoutb}
\end{figure}

\begin{figure}[ht!]
  \includegraphics[width=\linewidth]{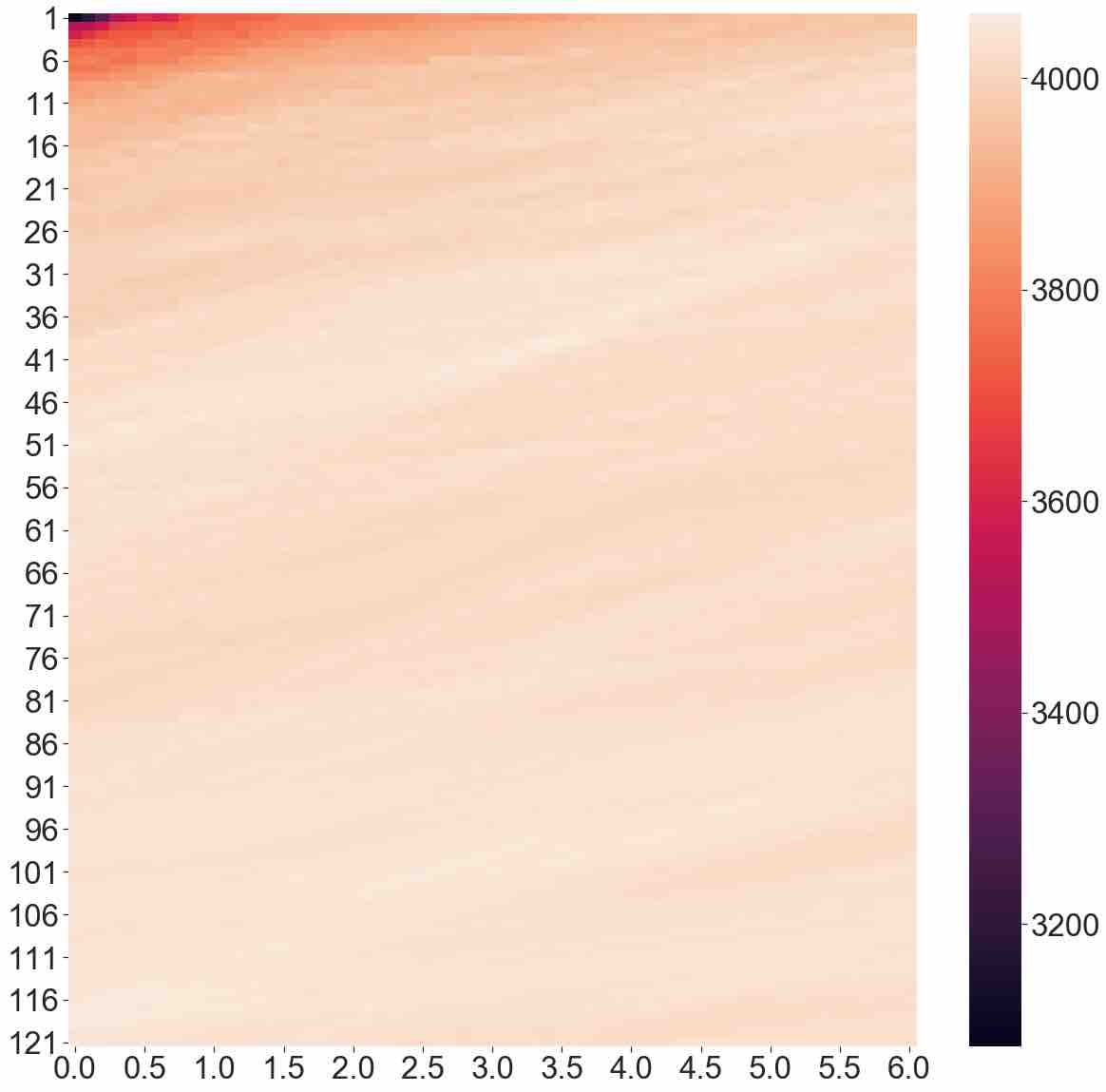}
  \caption{Benefit percentage attained AUCBE (optimal values for $\beta_1$ and $\beta_2$ are 39 and 3.5 respectively) \textbf{up to budget 45} on \textbf{v11.1}}
  \label{fig:v1145}
\end{figure}

\begin{figure}[ht!]
  \includegraphics[width=\linewidth]{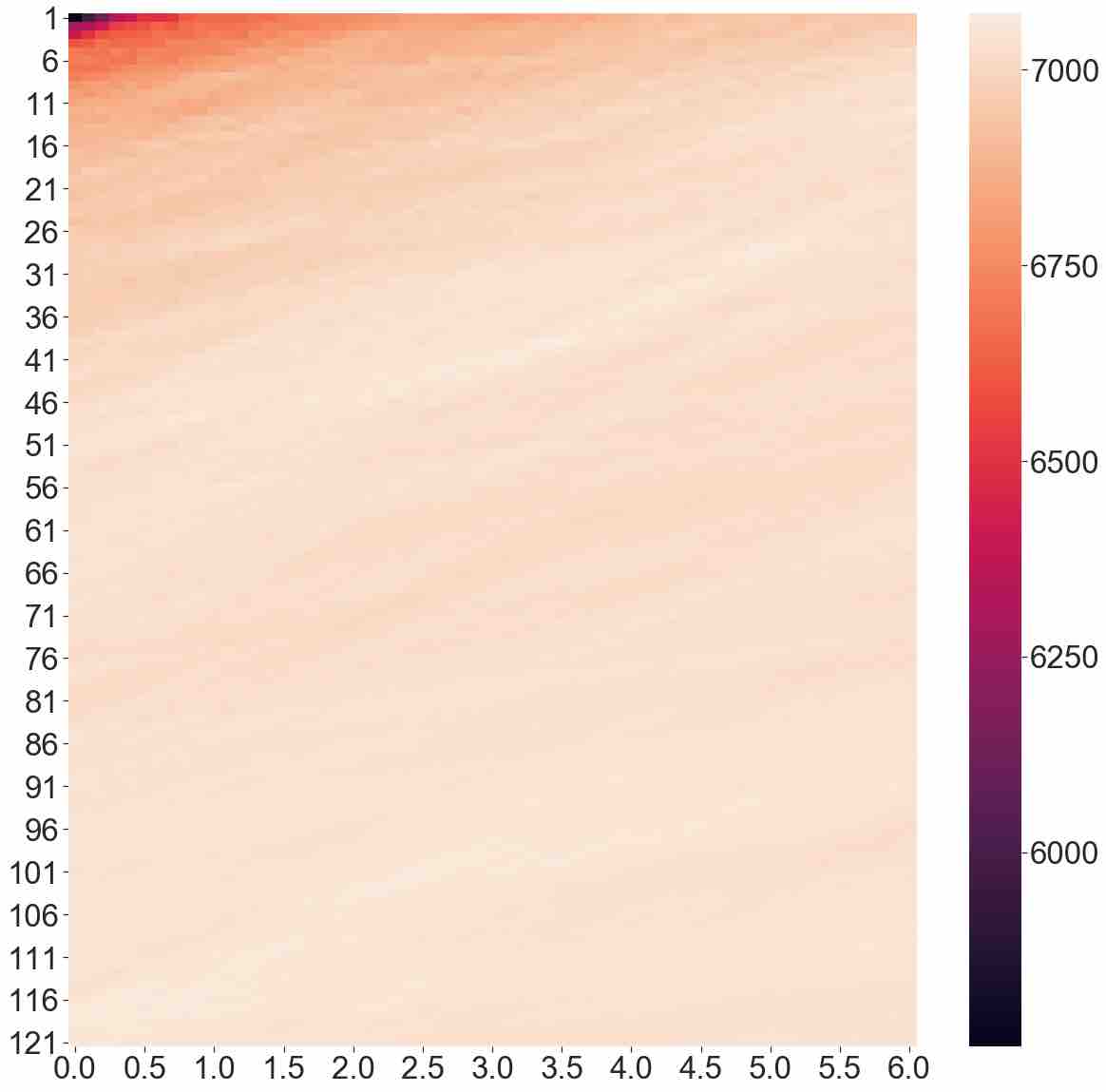}
  \caption{Benefit percentage attained AUCBE (optimal values for $\beta_1$ and $\beta_2$ are 39 and 3.5 respectively) \textbf{up to budget 65} on \textbf{v11.1}}
  \label{fig:v1165}
\end{figure}

\begin{figure}[ht!]
  \includegraphics[width=\linewidth]{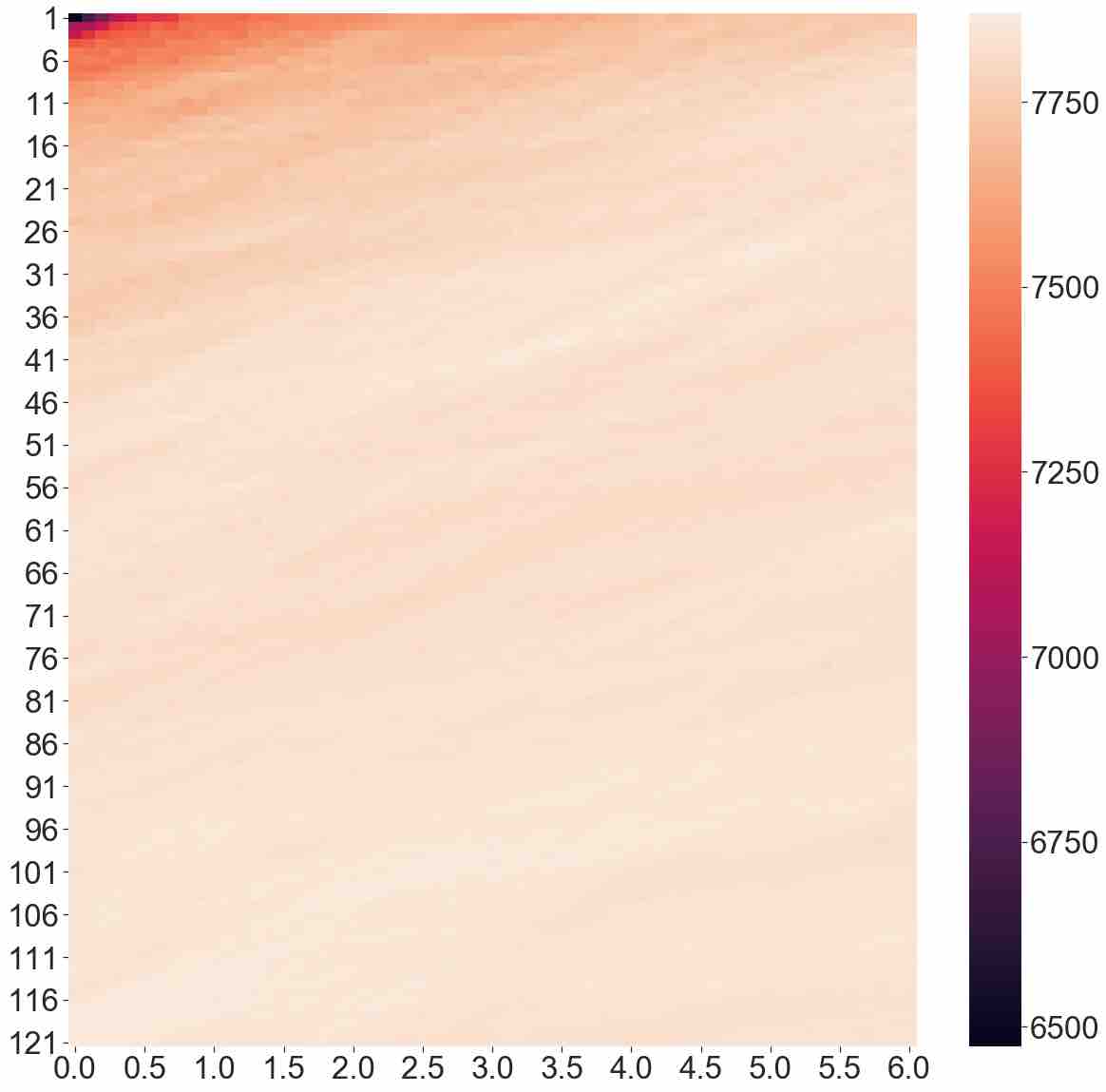}
  \caption{Benefit percentage attained AUCBE (optimal values for $\beta_1$ and $\beta_2$ are 103 and 2.2 respectively) \textbf{up to budget 70} on \textbf{v11.1}}
  \label{fig:v1170}
\end{figure}

\begin{figure}[ht!]
  \includegraphics[width=\linewidth]{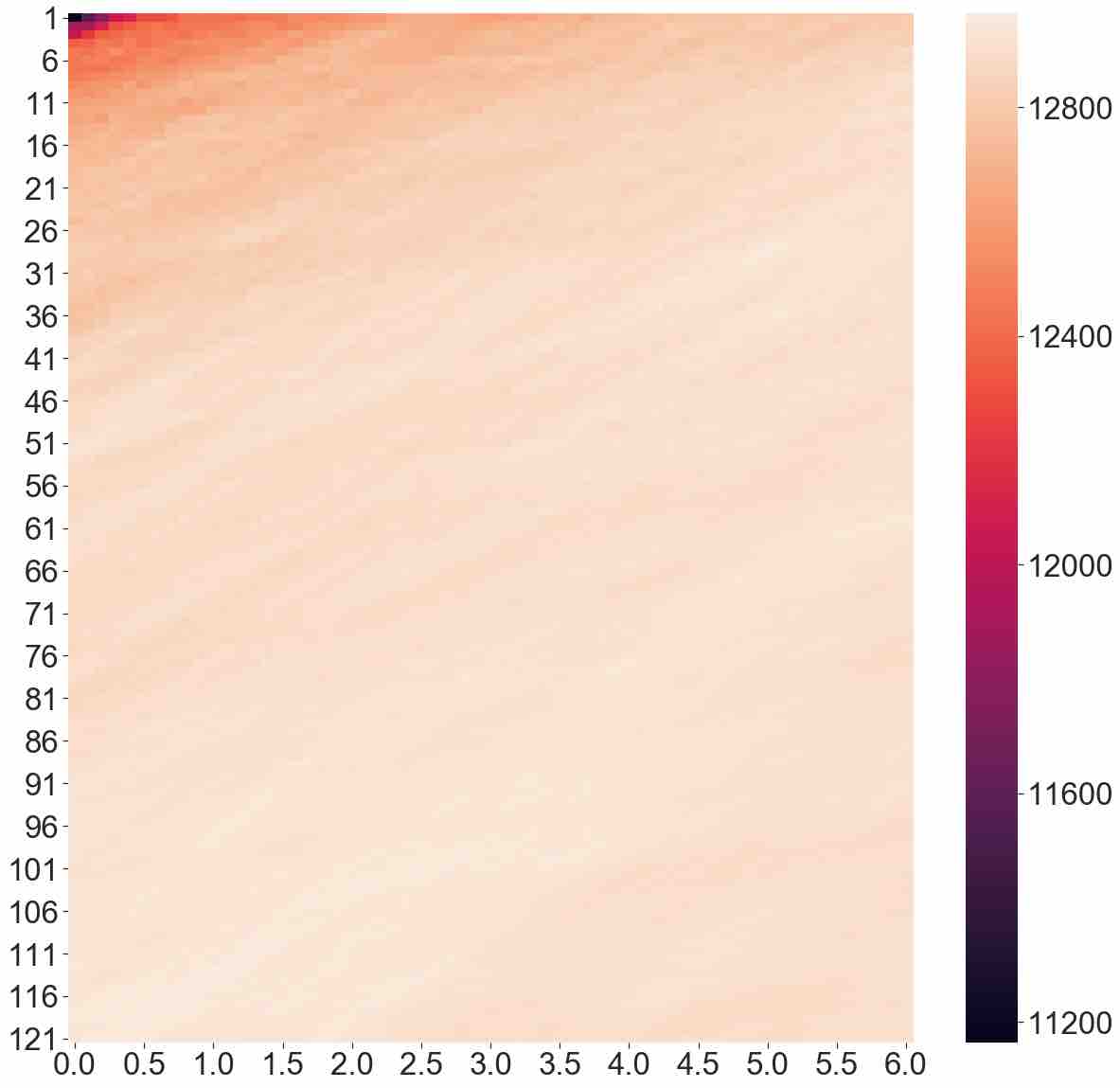}
  \caption{Benefit percentage attained AUCBE (optimal values for $\beta_1$ and $\beta_2$ are 118 and 0.6 respectively) \textbf{up to budget 100} on \textbf{v11.1}}
  \label{fig:v11100}
\end{figure}

\begin{figure}[ht!]
  \includegraphics[width=\linewidth]{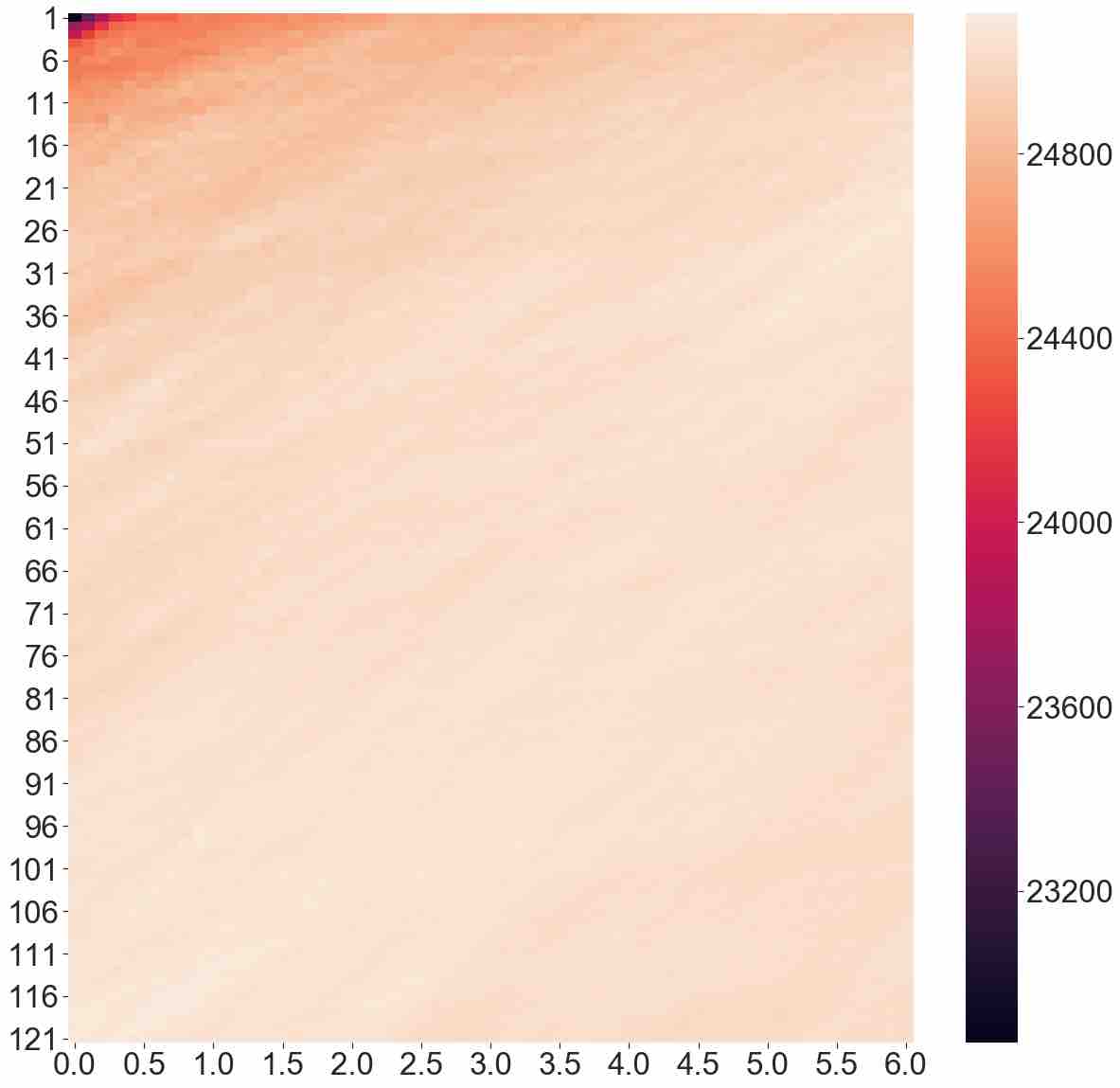}
  \caption{Benefit percentage attained AUCBE (optimal values for $\beta_1$ and $\beta_2$ are 118 and 0.6 respectively) \textbf{without budget limitation} on \textbf{11.1}}
  \label{fig:v11withoutb}
\end{figure}

For each budget limit and each dataset we performed hyper-parameter optimization ($\beta_1$, and $\beta_2$) using Myopic approach. For each dataset and different budget limit, there is a heatmap (see \cref{fig:v645,fig:v665,fig:v670,fig:v6100,fig:v6withoutb,fig:v1045,fig:v1065,fig:v1070,fig:v10100,fig:v10withoutb,fig:v1145,fig:v1165,fig:v1170,fig:v11100,fig:v11withoutb}). Each cell of heatmap shows AUCBE of average percentage of benefit attained up to the budget limit. $\beta_1$ variable ranges from 1 to 130, and $\beta_2$ ranges from 0 to 6 with 0.1 intervals. For the sake of better visualization, heatmaps are cropped at y-axis ($\beta_1$) ranges 1 to 61, 1 to 101, and 1 to 121 for datasets v6.3, v10.1, and v11.3 respectively. For all heatmaps the y-axis shows $\beta_1$ and x-axis shows $\beta_2$.




%% file: plots/benefit_cost_budget_45_v10.tex
\pgfplotstableread[col sep=comma,]{data/results/result_benefit_v10_29.csv}\ResultsSecondVersion

\begin{figure}[ht!]
\begin{tikzpicture}
	\begin{axis}[font=\small,
	    xmin=0,xmax=45,ymin=0,ymax=250,
		xlabel=Cumulative Effort Cost,
        ylabel=Cumulative Benefit Obtained,
            width=\linewidth,height = 5.8cm, 
        no markers,
        legend pos=south east,
		legend style={at={(0.50,-0.22)},anchor=north}
		]

		      \addplot [semithick, color = red ]table [x=Budget, y=DISCLOSE_Benefit_45] {\ResultsSecondVersion};
      \addlegendentry{DISCLOSE-v10.1}
		      \addplot [semithick, color = green ]table [x=Budget, y=Static_Benefit_45] {\ResultsSecondVersion};
      \addlegendentry{Static-v10.1}
	      \addplot [semithick,color = blue]table [x=Budget, y=MCTS_Benefit_45] {\ResultsSecondVersion};
      \addlegendentry{MCTS-v10.1}

\addplot [semithick,dotted, color = red ]table [x=Budget, y=DISCLOSE_45_0.25] {\ResultsSecondVersion};
		      \addplot [semithick, dotted,color = green ]table [x=Budget, y=Static_45_0.25] {\ResultsSecondVersion};
	      \addplot [semithick,dotted,color = blue ]table [x=Budget, y=MCTS_45_0.25] {\ResultsSecondVersion};

      \addplot [semithick, dashed,color = red ]table [x=Budget, y=DISCLOSE_45_0.75] {\ResultsSecondVersion};
		      \addplot [semithick,dashed, color = green ]table [x=Budget, y=Static_45_0.75] {\ResultsSecondVersion};
	      \addplot [semithick,dashed,color = blue ]table [x=Budget, y=MCTS_45_0.75] {\ResultsSecondVersion};

	\end{axis}
\end{tikzpicture}
		\caption{Cumulative benefit obtained as a function of cumulative effort cost (\textbf{up to budget 45}) on \textbf{v10.1}.}
	\label{fig:BenefitCost45v10}
\end{figure}

%% file: plots/benefit_cost_budget_45_v11.tex
\pgfplotstableread[col sep=comma,]{data/results/result_benefit_v11_29.csv}\ResultsThirdVersion

\begin{figure}[t]
\begin{tikzpicture}
	\begin{axis}[font=\small,legend style={font=\tiny},
	    xmin=0,xmax=45,ymin=0,ymax=180,
		xlabel=Cumulative Effort Cost,
        ylabel=Cumulative Benefit Obtained,
            width=\linewidth,height = 5.8cm, 
        no markers,
        legend pos=south east,
		legend style={at={(0.15,1)},anchor=north}
		]

     \addplot [semithick, color = red ]table [x=Budget, y=DISCLOSE_Benefit_45] {\ResultsThirdVersion};
      \addlegendentry{DISCLOSE}
    
		      \addplot [semithick, color = green ]table [x=Budget, y=Static_Benefit_45] {\ResultsThirdVersion};
      \addlegendentry{Static}
	      \addplot [semithick,color = blue ]table [x=Budget, y=MCTS_Benefit_45] {\ResultsThirdVersion};
      \addlegendentry{MCTS}

\addplot [semithick,dotted, color = red ]table [x=Budget, y=DISCLOSE_45_0.25] {\ResultsThirdVersion};
		      \addplot [semithick, dotted,color = green ]table [x=Budget, y=Static_45_0.25] {\ResultsThirdVersion};
	      \addplot [semithick,dotted,color = blue ]table [x=Budget, y=MCTS_45_0.25] {\ResultsThirdVersion};

      \addplot [semithick, dashed,color = red ]table [x=Budget, y=DISCLOSE_45_0.75] {\ResultsThirdVersion};
		      \addplot [semithick,dashed, color = green ]table [x=Budget, y=Static_45_0.75] {\ResultsThirdVersion};
	      \addplot [semithick,dashed,color = blue ]table [x=Budget, y=MCTS_45_0.75] {\ResultsThirdVersion};
	\end{axis}
\end{tikzpicture}
		\caption{Cumulative benefit obtained as a function of cumulative effort cost (\textbf{up to budget 45}) on~\textbf{v11.3}.}
	\label{fig:BenefitCost45v11}
\end{figure}

%% file: plots/benefit_cost_budget_65_v10.tex
\pgfplotstableread[col sep=comma,]{data/results/result_benefit_v10_29.csv}\ResultsSecondVersion

\begin{figure}[ht!]
\begin{tikzpicture}
	\begin{axis}[font=\small,
	    xmin=0,xmax=65,ymin=0,ymax=250,
		xlabel=Cumulative Effort Cost,
        ylabel=Cumulative Benefit Obtained,
            width=\linewidth,height = 5.8cm, 
        no markers,
        legend pos=south east,
		legend style={at={(0.50,-0.22)},anchor=north}
		]

		      \addplot [semithick, color = red ]table [x=Budget, y=DISCLOSE_Benefit_65] {\ResultsSecondVersion};
      \addlegendentry{DISCLOSE-v10.1}
		      \addplot [semithick, color = green ]table [x=Budget, y=Static_Benefit_65] {\ResultsSecondVersion};
      \addlegendentry{Static-v10.1}
	      \addplot [semithick,color = blue]table [x=Budget, y=MCTS_Benefit_65] {\ResultsSecondVersion};
      \addlegendentry{MCTS-v10.1}

\addplot [semithick,dotted, color = red ]table [x=Budget, y=DISCLOSE_65_0.25] {\ResultsSecondVersion};
		      \addplot [semithick, dotted,color = green ]table [x=Budget, y=Static_65_0.25] {\ResultsSecondVersion};
	      \addplot [semithick,dotted,color = blue ]table [x=Budget, y=MCTS_65_0.25] {\ResultsSecondVersion};

      \addplot [semithick, dashed,color = red ]table [x=Budget, y=DISCLOSE_65_0.75] {\ResultsSecondVersion};
		      \addplot [semithick,dashed, color = green ]table [x=Budget, y=Static_65_0.75] {\ResultsSecondVersion};
	      \addplot [semithick,dashed,color = blue ]table [x=Budget, y=MCTS_65_0.75] {\ResultsSecondVersion};

	\end{axis}
\end{tikzpicture}
		\caption{Cumulative benefit obtained as a function of cumulative effort cost (\textbf{up to budget 65}) on \textbf{v10.1}.}
	\label{fig:BenefitCost65v10}
\end{figure}

%% file: plots/benefit_cost_budget_65_v11.tex
\pgfplotstableread[col sep=comma,]{data/results/result_benefit_v11_29.csv}\ResultsThirdVersion

\begin{figure}[t]
\begin{tikzpicture}
	\begin{axis}[font=\small,legend style={font=\tiny},
	    xmin=0,xmax=65,ymin=0,ymax=180,
		xlabel=Cumulative Effort Cost,
        ylabel=Cumulative Benefit Obtained,
            width=\linewidth,height = 5.8cm, 
        no markers,
        legend pos=south east,
		legend style={at={(0.15,1)},anchor=north}
		]

     \addplot [semithick, color = red ]table [x=Budget, y=DISCLOSE_Benefit_65] {\ResultsThirdVersion};
      \addlegendentry{DISCLOSE}
    
		      \addplot [semithick, color = green ]table [x=Budget, y=Static_Benefit_65] {\ResultsThirdVersion};
      \addlegendentry{Static}
	      \addplot [semithick,color = blue ]table [x=Budget, y=MCTS_Benefit_65] {\ResultsThirdVersion};
      \addlegendentry{MCTS}

\addplot [semithick,dotted, color = red ]table [x=Budget, y=DISCLOSE_65_0.25] {\ResultsThirdVersion};
		      \addplot [semithick, dotted,color = green ]table [x=Budget, y=Static_65_0.25] {\ResultsThirdVersion};
	      \addplot [semithick,dotted,color = blue ]table [x=Budget, y=MCTS_65_0.25] {\ResultsThirdVersion};

      \addplot [semithick, dashed,color = red ]table [x=Budget, y=DISCLOSE_65_0.75] {\ResultsThirdVersion};
		      \addplot [semithick,dashed, color = green ]table [x=Budget, y=Static_65_0.75] {\ResultsThirdVersion};
	      \addplot [semithick,dashed,color = blue ]table [x=Budget, y=MCTS_65_0.75] {\ResultsThirdVersion};
	\end{axis}
\end{tikzpicture}
		\caption{Cumulative benefit obtained as a function of cumulative effort cost (\textbf{up to budget 65}) on~\textbf{v11.3}.}
	\label{fig:BenefitCost65v11}
\end{figure}

%% file: plots/benefit_cost_budget_70_v6.tex
\pgfplotstableread[col sep=comma,]{data/results/result_benefit_v6_31.csv}\ResultsFirstVersion

\begin{figure}[t]
\begin{tikzpicture}
	\begin{axis}[font=\small,
	    xmin=0,xmax=70,ymin=0,ymax=250,
		xlabel=Cumulative Effort Cost,
        ylabel=Cumulative Benefit Obtained,
            width=\linewidth,height = 5.8cm, 
        no markers,
        legend pos=south east,
		legend style={at={(0.50,-0.22)},anchor=north}
		]

		      \addplot [semithick, color = red ]table [x=Budget, y=DISCLOSE_Benefit_70] {\ResultsFirstVersion};
      \addlegendentry{DISCLOSE-v6.3}
		      \addplot [semithick, color = green ]table [x=Budget, y=Static_Benefit_70] {\ResultsFirstVersion};
      \addlegendentry{Static-v6.3}
	      \addplot [semithick,color = blue]table [x=Budget, y=MCTS_Benefit_70] {\ResultsFirstVersion};
      \addlegendentry{MCTS-v6.3}

\addplot [semithick,dotted, color = red ]table [x=Budget, y=DISCLOSE_70_0.25] {\ResultsFirstVersion};
		      \addplot [semithick, dotted,color = green ]table [x=Budget, y=Static_70_0.25] {\ResultsFirstVersion};
	      \addplot [semithick,dotted,color = blue ]table [x=Budget, y=MCTS_70_0.25] {\ResultsFirstVersion};

      \addplot [semithick, dashed,color = red ]table [x=Budget, y=DISCLOSE_70_0.75] {\ResultsFirstVersion};
		      \addplot [semithick,dashed, color = green ]table [x=Budget, y=Static_70_0.75] {\ResultsFirstVersion};
	      \addplot [semithick,dashed,color = blue ]table [x=Budget, y=MCTS_70_0.75] {\ResultsFirstVersion};

	\end{axis}
\end{tikzpicture}
		\caption{Cumulative benefit obtained as a function of cumulative effort cost (\textbf{up to budget 70}) on \textbf{v6.3}.}
	\label{fig:BenefitCost70v6}
\end{figure}

%% file: plots/benefit_cost_budget_70_v10.tex
\pgfplotstableread[col sep=comma,]{data/results/result_benefit_v10_29.csv}\ResultsSecondVersion

\begin{figure}[t]
\begin{tikzpicture}
	\begin{axis}[font=\small,
	    xmin=0,xmax=70,ymin=0,ymax=250,
		xlabel=Cumulative Effort Cost,
        ylabel=Cumulative Benefit Obtained,
            width=\linewidth,height = 5.8cm, 
        no markers,
        legend pos=south east,
		legend style={at={(0.50,-0.22)},anchor=north}
		]

		      \addplot [semithick, color = red ]table [x=Budget, y=DISCLOSE_Benefit_70] {\ResultsSecondVersion};
      \addlegendentry{DISCLOSE-v10.1}
		      \addplot [semithick, color = green ]table [x=Budget, y=Static_Benefit_70] {\ResultsSecondVersion};
      \addlegendentry{Static-v10.1}
	      \addplot [semithick,color = blue]table [x=Budget, y=MCTS_Benefit_70] {\ResultsSecondVersion};
      \addlegendentry{MCTS-v10.1}

\addplot [semithick,dotted, color = red ]table [x=Budget, y=DISCLOSE_70_0.25] {\ResultsSecondVersion};
		      \addplot [semithick, dotted,color = green ]table [x=Budget, y=Static_70_0.25] {\ResultsSecondVersion};
	      \addplot [semithick,dotted,color = blue ]table [x=Budget, y=MCTS_70_0.25] {\ResultsSecondVersion};

      \addplot [semithick, dashed,color = red ]table [x=Budget, y=DISCLOSE_70_0.75] {\ResultsSecondVersion};
		      \addplot [semithick,dashed, color = green ]table [x=Budget, y=Static_70_0.75] {\ResultsSecondVersion};
	      \addplot [semithick,dashed,color = blue ]table [x=Budget, y=MCTS_70_0.75] {\ResultsSecondVersion};

	\end{axis}
\end{tikzpicture}
		\caption{Cumulative benefit obtained  as a function of cumulative effort cost (\textbf{up to budget 70}) on \textbf{v10.1}.}
	\label{fig:BenefitCost70v10}
\end{figure}

%% file: plots/benefit_cost_budget_70_v11.tex
\pgfplotstableread[col sep=comma,]{data/results/result_benefit_v11_29.csv}\ResultsThirdVersion

\begin{figure}[t]
\begin{tikzpicture}
	\begin{axis}[font=\small,
	    xmin=0,xmax=70,ymin=0,ymax=250,
		xlabel=Cumulative Effort Cost,
        ylabel=Cumulative Benefit Obtained,
            width=\linewidth,height = 5.8cm, 
        no markers,
        legend pos=south east,
		legend style={at={(0.50,-0.22)},anchor=north}
		]

		      \addplot [semithick, color = red ]table [x=Budget, y=DISCLOSE_Benefit_70] {\ResultsThirdVersion};
      \addlegendentry{DISCLOSE-v11.3}
		      \addplot [semithick, color = green ]table [x=Budget, y=Static_Benefit_70] {\ResultsThirdVersion};
      \addlegendentry{Static-v11.3}
	      \addplot [semithick,color = blue]table [x=Budget, y=MCTS_Benefit_70] {\ResultsThirdVersion};
      \addlegendentry{MCTS-v11.3}

\addplot [semithick,dotted, color = red ]table [x=Budget, y=DISCLOSE_70_0.25] {\ResultsThirdVersion};
		      \addplot [semithick, dotted,color = green ]table [x=Budget, y=Static_70_0.25] {\ResultsThirdVersion};
	      \addplot [semithick,dotted,color = blue ]table [x=Budget, y=MCTS_70_0.25] {\ResultsThirdVersion};

      \addplot [semithick, dashed,color = red ]table [x=Budget, y=DISCLOSE_70_0.75] {\ResultsThirdVersion};
		      \addplot [semithick,dashed, color = green ]table [x=Budget, y=Static_70_0.75] {\ResultsThirdVersion};
	      \addplot [semithick,dashed,color = blue ]table [x=Budget, y=MCTS_70_0.75] {\ResultsThirdVersion};

	\end{axis}
\end{tikzpicture}
		\caption{Cumulative benefit obtained as a function of cumulative effort cost (\textbf{up to budget 70}) on \textbf{v11.3}.}
	\label{fig:BenefitCost70v11}
\end{figure}

%% file: plots/benefit_cost_budget_100_v6.tex
\pgfplotstableread[col sep=comma,]{data/results/result_benefit_v6_31.csv}\ResultsFirstVersion

\begin{figure}[t]
\begin{tikzpicture}
	\begin{axis}[font=\small,
	    xmin=0,xmax=101,ymin=0,ymax=250,
		xlabel=Cumulative Effort Cost,
        ylabel=Cumulative Benefit Obtained,
            width=\linewidth,height = 5.8cm, 
        no markers,
        legend pos=south east,
		legend style={at={(0.50,-0.22)},anchor=north}
		]

		      \addplot [semithick, color = red ]table [x=Budget, y=DISCLOSE_Benefit_100] {\ResultsFirstVersion};
      \addlegendentry{DISCLOSE-v6.3}
		      \addplot [semithick, color = green ]table [x=Budget, y=Static_Benefit_100] {\ResultsFirstVersion};
      \addlegendentry{Static-v6.3}
	      \addplot [semithick,color = blue]table [x=Budget, y=MCTS_Benefit_100] {\ResultsFirstVersion};
      \addlegendentry{MCTS-v6.3}

\addplot [semithick,dotted, color = red ]table [x=Budget, y=DISCLOSE_100_0.25] {\ResultsFirstVersion};
		      \addplot [semithick, dotted,color = green ]table [x=Budget, y=Static_100_0.25] {\ResultsFirstVersion};
	      \addplot [semithick,dotted,color = blue ]table [x=Budget, y=MCTS_100_0.25] {\ResultsFirstVersion};

      \addplot [semithick, dashed,color = red ]table [x=Budget, y=DISCLOSE_100_0.75] {\ResultsFirstVersion};
		      \addplot [semithick,dashed, color = green ]table [x=Budget, y=Static_100_0.75] {\ResultsFirstVersion};
	      \addplot [semithick,dashed,color = blue ]table [x=Budget, y=MCTS_100_0.75] {\ResultsFirstVersion};

	\end{axis}
\end{tikzpicture}
		\caption{Cumulative benefit obtained as a function of cumulative effort cost (\textbf{up to budget 100}) on \textbf{v6.3}.}
	\label{fig:BenefitCost100v6}
\end{figure}

%% file: plots/benefit_cost_budget_100_v10.tex
\pgfplotstableread[col sep=comma,]{data/results/result_benefit_v10_29.csv}\ResultsSecondVersion

\begin{figure}[t]
\begin{tikzpicture}
	\begin{axis}[font=\small,
	    xmin=0,xmax=101,ymin=0,ymax=250,
		xlabel=Cumulative Effort Cost,
        ylabel=Cumulative Benefit Obtained,
            width=\linewidth,height = 5.8cm, 
        no markers,
        legend pos=south east,
		legend style={at={(0.50,-0.22)},anchor=north}
		]

		      \addplot [semithick, color = red ]table [x=Budget, y=DISCLOSE_Benefit_100] {\ResultsSecondVersion};
      \addlegendentry{DISCLOSE-v10.1}
		      \addplot [semithick, color = green ]table [x=Budget, y=Static_Benefit_100] {\ResultsSecondVersion};
      \addlegendentry{Static-v10.1}
	      \addplot [semithick,color = blue]table [x=Budget, y=MCTS_Benefit_100] {\ResultsSecondVersion};
      \addlegendentry{MCTS-v10.1}

\addplot [semithick,dotted, color = red ]table [x=Budget, y=DISCLOSE_100_0.25] {\ResultsSecondVersion};
		      \addplot [semithick, dotted,color = green ]table [x=Budget, y=Static_100_0.25] {\ResultsSecondVersion};
	      \addplot [semithick,dotted,color = blue ]table [x=Budget, y=MCTS_100_0.25] {\ResultsSecondVersion};

      \addplot [semithick, dashed,color = red ]table [x=Budget, y=DISCLOSE_100_0.75] {\ResultsSecondVersion};
		      \addplot [semithick,dashed, color = green ]table [x=Budget, y=Static_100_0.75] {\ResultsSecondVersion};
	      \addplot [semithick,dashed,color = blue ]table [x=Budget, y=MCTS_100_0.75] {\ResultsSecondVersion};

	\end{axis}
\end{tikzpicture}
		\caption{Cumulative benefit obtained   as a function of cumulative effort cost (\textbf{up to budget 100}) on \textbf{v10.1}.}
	\label{fig:BenefitCost100v10}
\end{figure}

%% file: plots/benefit_cost_budget_100_v11.tex
\pgfplotstableread[col sep=comma,]{data/results/result_benefit_v11_29.csv}\ResultsThirdVersion

\begin{figure}[t]
\begin{tikzpicture}
	\begin{axis}[font=\small,
	    xmin=0,xmax=101,ymin=0,ymax=250,
		xlabel=Cumulative Effort Cost,
        ylabel=Cumulative Benefit Obtained,
            width=\linewidth,height = 5.8cm, 
        no markers,
        legend pos=south east,
		legend style={at={(0.50,-0.22)},anchor=north}
		]

		      \addplot [semithick, color = red ]table [x=Budget, y=DISCLOSE_Benefit_100] {\ResultsThirdVersion};
      \addlegendentry{DISCLOSE-v11.3}
		      \addplot [semithick, color = green ]table [x=Budget, y=Static_Benefit_100] {\ResultsThirdVersion};
      \addlegendentry{Static-v11.3}
	      \addplot [semithick,color = blue]table [x=Budget, y=MCTS_Benefit_100] {\ResultsThirdVersion};
      \addlegendentry{MCTS-v11.3}

\addplot [semithick,dotted, color = red ]table [x=Budget, y=DISCLOSE_100_0.25] {\ResultsThirdVersion};
		      \addplot [semithick, dotted,color = green ]table [x=Budget, y=Static_100_0.25] {\ResultsThirdVersion};
	      \addplot [semithick,dotted,color = blue ]table [x=Budget, y=MCTS_100_0.25] {\ResultsThirdVersion};

      \addplot [semithick, dashed,color = red ]table [x=Budget, y=DISCLOSE_100_0.75] {\ResultsThirdVersion};
		      \addplot [semithick,dashed, color = green ]table [x=Budget, y=Static_100_0.75] {\ResultsThirdVersion};
	      \addplot [semithick,dashed,color = blue ]table [x=Budget, y=MCTS_100_0.75] {\ResultsThirdVersion};

	\end{axis}
\end{tikzpicture}
		\caption{Cumulative benefit obtained  as a function of cumulative effort cost (\textbf{up to budget 100}) on \textbf{v11.3}.}
	\label{fig:BenefitCost100v11}
\end{figure}

%% file: plots/benefit_cost_without_budget_v6.tex
\pgfplotstableread[col sep=comma,]{data/results/result_benefit_v6_31.csv}\ResultsFirstVersion

\begin{figure}[t]
\begin{tikzpicture}
	\begin{axis}[font=\small,
	    xmin=0,xmax=174,ymin=0,ymax=250,
		xlabel=Cumulative Effort Cost,
        ylabel=Cumulative Benefit Obtained,
            width=\linewidth,height = 5.8cm, 
        no markers,
        legend pos=south east,
		legend style={at={(0.50,-0.22)},anchor=north}
		]

		      \addplot [semithick, color = red ]table [x=Budget, y=DISCLOSE_Benefit_withoutBudget] {\ResultsFirstVersion};
      \addlegendentry{DISCLOSE-v6.3}
		      \addplot [semithick, color = green ]table [x=Budget, y=Static_Benefit_withoutBudget] {\ResultsFirstVersion};
      \addlegendentry{Static-v6.3}
	      \addplot [semithick,color = blue]table [x=Budget, y=MCTS_Benefit_withoutBudget] {\ResultsFirstVersion};
      \addlegendentry{MCTS-v6.3}

\addplot [semithick,dotted, color = red ]table [x=Budget, y=DISCLOSE_183_0.25] {\ResultsFirstVersion};
		      \addplot [semithick, dotted,color = green ]table [x=Budget, y=Static_183_0.25] {\ResultsFirstVersion};
	      \addplot [semithick,dotted,color = blue ]table [x=Budget, y=MCTS_183_0.25] {\ResultsFirstVersion};

      \addplot [semithick, dashed,color = red ]table [x=Budget, y=DISCLOSE_183_0.75] {\ResultsFirstVersion};
		      \addplot [semithick,dashed, color = green ]table [x=Budget, y=Static_183_0.75] {\ResultsFirstVersion};
	      \addplot [semithick,dashed,color = blue ]table [x=Budget, y=MCTS_183_0.75] {\ResultsFirstVersion};

	\end{axis}
\end{tikzpicture}
		\caption{Cumulative benefit obtained  as a function of cumulative effort cost (\textbf{without budget limitation}) on \textbf{v6.3}.}
	\label{fig:BenefitCostwithoutbudgetv6}
\end{figure}

%% file: plots/benefit_cost_without_budget_v10.tex
\pgfplotstableread[col sep=comma,]{data/results/result_benefit_v10_29.csv}\ResultsSecondVersion

\begin{figure}[t]
\begin{tikzpicture}
	\begin{axis}[font=\small,
	    xmin=0,xmax=162,ymin=0,ymax=250,
		xlabel=Cumulative Effort Cost,
        ylabel=Cumulative Benefit Obtained,
            width=\linewidth,height = 5.8cm, 
        no markers,
        legend pos=south east,
		legend style={at={(0.50,-0.22)},anchor=north}
		]

		      \addplot [semithick, color = red ]table [x=Budget, y=DISCLOSE_Benefit_withoutBudget] {\ResultsSecondVersion};
      \addlegendentry{DISCLOSE-v10.1}
		      \addplot [semithick, color = green ]table [x=Budget, y=Static_Benefit_withoutBudget] {\ResultsSecondVersion};
      \addlegendentry{Static-v10.1}
	      \addplot [semithick,color = blue]table [x=Budget, y=MCTS_Benefit_withoutBudget] {\ResultsSecondVersion};
      \addlegendentry{MCTS-v10.1}

\addplot [semithick,dotted, color = red ]table [x=Budget, y=DISCLOSE_171_0.25] {\ResultsSecondVersion};
		      \addplot [semithick,color = green ]table [x=Budget, y=Static_171_0.25] {\ResultsSecondVersion};
	      \addplot [semithick,color = blue ]table [x=Budget, y=MCTS_171_0.25] {\ResultsSecondVersion};

      \addplot [semithick, dashed,color = red ]table [x=Budget, y=DISCLOSE_171_0.75] {\ResultsSecondVersion};
		      \addplot [semithick,dashed, color = green ]table [x=Budget, y=Static_171_0.75] {\ResultsSecondVersion};
	      \addplot [semithick,dashed,color = blue ]table [x=Budget, y=MCTS_171_0.75] {\ResultsSecondVersion};

	\end{axis}
\end{tikzpicture}
		\caption{Cumulative benefit obtained  as a function of cumulative effort cost (\textbf{without budget limitation}) on \textbf{v10.1}.}
	\label{fig:BenefitCostwithoutbudgetv10}
\end{figure}

%% file: plots/benefit_cost_without_budget_v11.tex
\pgfplotstableread[col sep=comma,]{data/results/result_benefit_v11_29.csv}\ResultsThirdVersion

\begin{figure}[t]
\begin{tikzpicture}
	\begin{axis}[font=\small,
	    xmin=0,xmax=162,ymin=0,ymax=250,
		xlabel=Cumulative Effort Cost,
        ylabel=Cumulative Benefit Obtained,
            width=\linewidth,height = 5.8cm, 
        no markers,
        legend pos=south east,
		legend style={at={(0.50,-0.22)},anchor=north}
		]

		      \addplot [semithick, color = red ]table [x=Budget, y=DISCLOSE_Benefit_withoutBudget] {\ResultsThirdVersion};
      \addlegendentry{DISCLOSE-v11.3}
		      \addplot [semithick, color = green ]table [x=Budget, y=Static_Benefit_withoutBudget] {\ResultsThirdVersion};
      \addlegendentry{Static-v11.3}
	      \addplot [semithick,color = blue]table [x=Budget, y=MCTS_Benefit_withoutBudget] {\ResultsThirdVersion};
      \addlegendentry{MCTS-v11.3}

\addplot [semithick,dotted, color = red ]table [x=Budget, y=DISCLOSE_171_0.25] {\ResultsThirdVersion};
		      \addplot [semithick, dotted,color = green ]table [x=Budget, y=Static_171_0.25] {\ResultsThirdVersion};
	      \addplot [semithick,dotted,color = blue ]table [x=Budget, y=MCTS_171_0.25] {\ResultsThirdVersion};

      \addplot [semithick, dashed,color = red ]table [x=Budget, y=DISCLOSE_171_0.75] {\ResultsThirdVersion};
		      \addplot [semithick,dashed, color = green ]table [x=Budget, y=Static_171_0.75] {\ResultsThirdVersion};
	      \addplot [semithick,dashed,color = blue ]table [x=Budget, y=MCTS_171_0.75] {\ResultsThirdVersion};

	\end{axis}
\end{tikzpicture}
		\caption{Cumulative benefit obtained as a function of cumulative effort cost (\textbf{without budget limitation}) on \textbf{v11.3}.}
	\label{fig:BenefitCostwithoutbudgetv11}
\end{figure}